\documentclass[12pt]{article}
\pdfoutput=1
\usepackage[centertags]{amsmath}
\usepackage{amsfonts}
\usepackage{amssymb}
\usepackage{amsthm}
\usepackage{cite}
\usepackage{youngtab}
\usepackage{graphicx}
\usepackage{etoolbox}
\usepackage{multirow}
\usepackage{upgreek}

\newcommand{\beq}{\begin{equation}}
\newcommand{\eeq}[1]{\label{#1}\end{equation}}
\newcommand{\bea}{\begin{eqnarray}}
\newcommand{\eea}[1]{\label{#1}\end{eqnarray}}

\newcommand{\ba}{\begin{align}}
\newcommand{\ea}{\end{align}}

\newcommand{\marg}[1]{}

\newcommand{\rf}[1]{(\ref{#1})}

\renewcommand{\overleftarrow}[1]{\overset\leftarrow #1} 

\def\tf{\tfrac}
\def\de{\partial}
\def\ds{\displaystyle{\not{\!\partial\!\,}}}
\def\pss{\displaystyle{\not{\!\psi\!\,}}}
\def\Pss{\displaystyle{\not{\!\Psi\!\,}}}

\def\As{\displaystyle{\not{\;\!\!\!\!A}}}
\def\Bs{\displaystyle{\not{\;\!\!\!\!B}}}
\def\Cs{\displaystyle{\not{\;\!\!\!C}}}

\def\a{\alpha}
\def\b{\beta}
\def\g{\gamma}
\def\G{\Gamma}
\def\d{\delta}
\def\D{\Delta}
\def\e{\epsilon}
\def\ve{\varepsilon}

\def\h{\eta}

\def\Th{\Theta}

\def\k{\kappa}
\def\l{\lambda}
\def\m{\mu}
\def\n{\nu}
\def\x{\xi}

\def\r{\rho}

\def\s{\sigma}

\def\t{\tau}

\def\ps{\psi}
\def\Ps{\Psi}
\def\o{\omega}

\begin{document}
\numberwithin{equation}{section}
\setlength{\topmargin}{-1cm} \setlength{\oddsidemargin}{0cm}
\setlength{\evensidemargin}{0cm}

\begin{titlepage}
\begin{center}
{\Large \bf The Uniqueness of Hypergravity}

\vspace{20pt}

{\large Rakibur Rahman$^{\,a,b}$}

\vspace{12pt}
$^a$ Department of Physics, University of Dhaka, Dhaka 1000, Bangladesh\\
\vspace{6pt}
$^b$ Max-Planck-Institut f\"ur Gravitationsphysik (Albert-Einstein-Institut)\\
     Am M\"uhlenberg 1, D-14476 Potsdam-Golm, Germany\\
\end{center}
\vspace{20pt}

\begin{abstract}
We show that consistent interactions of a spin-2 and a higher-spin Majorana fermion gauge fields
in 3D flat space lead uniquely to Aragone-Deser hypergravity or its generalization.
Our analysis employs the BRST-cohomological techniques, and works in the metric-like formulation under
the assumptions of locality, parity and Poincar\'e invariance. Local hypersymmetry shows up as the unique
consistent deformation of the gauge transformations. An extension of the theory with fermion
flavors does not change these features, while a cosmological deformation becomes obstructed
in the absence of other degrees of freedom and/or non-locality.

\end{abstract}

\end{titlepage}

\newpage
\section{Introduction}\label{sec:Intro}

Higher-spin theories in three space-time dimensions have generated a lot of interests in recent years. The obstructions to minimal gravitational interaction of massless
higher-spin fields~\cite{Old,gpv,Aragone,ww,New} disappear in three dimensions where the Weyl tensor vanishes. In 3D flat space one can write down, among other things,
a consistent theory of a spin-$5/2$ gauge field minimally coupled to gravity$-$the well-known hypergravity theory of Aragone and Deser~\cite{Aragone:1983sz}. In itself a
higher-spin generalization of 3D supergravity, this theory is dictated by an extension of the Poincar\'e group$-$the hyper-Poincar\'e group$-$that includes spin-$3/2$
fermionic generators. This however does not contradict the Haag-{\L}opusza\'nski-Sohnius theorem~\cite{Haag:1974qh} that applies only to $D\geq4$. Indeed, one can
reformulate hypergravity in a way that incorporates the hyper-Poincar\'e group as local gauge symmetry~\cite{Fuentealba:2015jma}. To be more specific, hypergravity can
be described by the Chern-Simons action of a hyper-Poincar\'e-valued gauge field. These constructions can be generalized to an arbitrary-spin massless fermion
with an arbitrary number of flavors, minimally coupled to General Relativity.

In this article we investigate the uniqueness of (generalized) hypergravity as a consistent interacting gauge theory of a higher-spin fermion and gravity. For this purpose
we use the BRST-cohomological techniques based on antifield formalism~\cite{Barnich:1993vg,Henneaux:1997bm} under the assumptions of locality, parity and Poincar\'e invariance.
The same techniques have been employed in proving under certain reasonable assumptions the uniqueness of a number of physical theories: Yang-Mills theory~\cite{Barnich:1993pa}, General
Relativity~\cite{Boulanger:2000rq} and supergravity theories in $D\!=\!4$~\cite{Boulanger:2001wq,Boulanger:2018fei}. Along this line, we start
in three space-time dimensions with the free system of a massless arbitrary-spin Majorana fermion and a spin-2 gauge field. Nontrivial cubic and higher-order deformations
compatible with the gauge symmetries of this system can then be derived systematically. The question we seek to answer is whether the set of consistent deformations
of the free theory leads uniquely to generalized hypergravity. Our approach assumes neither general covariance nor local hypersymmetry to begin with;
they instead would follow automatically as a possibility, if not the unique one.

The organization of the article is as follows. The remaining of this section presents our main results and conventions.
Section~\ref{sec:A-D} is a brief exposition of generalized hypergravity.
Section~\ref{sec:BRST} introduces the much-needed
machinery of the BRST deformation scheme for irreducible gauge theories~\cite{Barnich:1993vg,Henneaux:1997bm}. Sections~\ref{sec:Metric} and~\ref{sec:GenH} constitute
the bulk of this paper; they use the metric-like formulation of higher spins to investigate respectively the uniqueness of Aragone-Deser hypergravity with a spin-$5/2$
Majorana fermion and that of its generalizations to an arbitrary-spin fermion with flavors. The consequences of a cosmological term are
studied in Section~\ref{sec:CC}, which sheds light on
the nature of (generalized) hypergravity in (Anti-)de Sitter space. Some remarks appear in Section~\ref{sec:remarks}. Three appendices are added in order to provide
the reader with some useful details.

\subsection*{Results}\label{subsec:results}

\begin{itemize}
  \item Assuming locality, parity and Poincar\'e invariance, consistent interactions of a spin-2 and a spin-$5/2$ Majorana gauge fields lead
  uniquely to Aragone-Deser hypergravity.
  \item The uniqueness continues to hold for arbitrary fermion spin $s=n+3/2$.
  As a byproduct, the uniqueness of three-dimensional supergravity is proven for $n=0$.
  \item Generalized hypergravity is consistent for an arbitrary number of fermion flavors.
  \item Local hypersymmetry and its higher-spin counterparts follow automatically as the unique consistent deformations of the gauge transformations.
  \item Another byproduct is the explicit demonstration that the cohomological obstruction to minimal gravitational coupling of higher-spin
  fermions disappears in $D=3$.
 \item There is no consistent local gauge theory of a higher-spin Majorana fermion and gravity in the presence of a cosmological term. Non-locality and/or
 additional degrees of freedom are required in (Anti-)de Sitter space.
\end{itemize}

\subsubsection*{Conventions \& Notations}\label{subsec:convnot}

We adopt the conventions of Ref.~\cite{Freedman:2012zz}, and work exclusively in $D\!=\!3$ with metric signature $(-++)$.
Fiber indices and world indices are denoted with lower case Roman letters and Greek letters respectively.
The $\g$-matrices satisfy the Clifford algebra: $\{\g^a,\g^b\}=+2\h^{ab}$, and $\g^{a\,\dagger}=\h^{aa}\g^a$.
Totally antisymmetric product of $\g$-matrices,
$\g^{a_1\cdots a_r}\equiv\g^{[a_1}\g^{a_2}\cdots\g^{a_r]}$, have a unit weight, where $[a_1\cdots a_r]$ denotes a totally antisymmetric
expression in the indices $a_1,\ldots,a_r$ with a normalization factor $\tf{1}{r!}$. The totally symmetric expression $(a_1\cdots a_r)$
carries the same normalization.
The Levi-Civita symbol is normalized as $\ve_{012}=+1$.

We exclusively deal with Majorana spinors. A Majorana spinor $\chi$ obeys: $\chi^C=\b\chi$, where the ``phase'' $\b$ is $+1$ $(-1)$ for a real (imaginary) spinor.
Majorana spinors $\chi_i$, $i=1,2$, with ``phase'' $\b_i$ and Grassmann parity $\e_i$, follow the bilinear identity:
\beq \bar\chi_1\g^{a_1\cdots a_r}\chi_2=(-)^{1+\e_1\e_2}\left(\b_1\b_2\right)t_r\,\bar\chi_2\g^{a_1\cdots a_r}\chi_1,\nonumber\eeq{}
where a ``bar'' denotes Majorana conjugation, and $t_r=+1\,(-1)$ for $r=0,3\,(1,2)$.

A ``slash'' denotes a contraction with $\gamma$-matrix, e.g., $\displaystyle{\not{\!\!A\!\,}}=\g^a A_a$, whereas
a ``prime'' denotes a trace w.r.t.~Minkowski metric, e.g., $h^{\prime}=\h^{\m\n}h_{\m\n}=h^{\m}_{~\m}$.
Finally, the symbol ``$\doteq$'' stands for equality of expressions up to a total derivative.

\section{Generalized Hypergravity}\label{sec:A-D}

This section gives a brief account of generalized hypergravity. First, the frame-like version of the theory is presented. Then, upon integrating out the spin connection, the metric-like version of the theory is obtained.

\subsection{Frame-Like Version}\label{sec:F-H}

Generalized hypergravity is the theory of a massless Majorana fermion $\uppsi_\m{}^{a_1\cdots a_n}$, of spin $s=n+3/2$, minimally coupled of General Relativity. The fermion field
is a dreibein-like gauge fermion$-$completely symmetric and $\g$-traceless in the frame-like indices~\cite{Aragone:1983sz,V-flat,Aragone:1980rk,Rahman:2017cxk}.
Here we adopt the notation of~\cite{Rahman:2015pzl} to use the short-hand symbol $\uppsi_\m{}^{a(n)}$ for it.

We consider a straightforward generalization of the theory presented in~\cite{Fuentealba:2015jma} by including a flavor index $I$ to the fermion, where $I=1,2,\ldots,N$.
The Chern-Simons generalized hypergravity action can be written, up to a boundary term, as:
\beq S_{H}=\frac{2}{\k^2}\int \left( 2e_aR^a-\bar\uppsi_{a(n),\,I}D\uppsi^{a(n),\,I}\right),\eeq{H1}
where $e^a=e_\m^a\,dx^\m$ is the dreibein 1-form, and $R^a$ the dualized curvature 2-form:
\beq R^a=D\o^a=d\o^a+\tfrac{1}{2}\ve^{abc}\o_b\o_c,\eeq{H2.1}
with $\o^a=\tfrac{1}{2}\ve^{abc}\o_{\m bc}\,dx^\m$ the dualized spin connection 1-form, while $\uppsi^{a(n),\,I}=\uppsi_\m{}^{a(n),\,I}dx^\m$ is the 1-form
corresponding to the fermion, with its covariant-derivative 2-form given by:
\beq D\uppsi^{a(n),\,I}=d\uppsi^{a(n),\,I}+\left(n+\tfrac{1}{2}\right)\o_b\g^b\uppsi^{a(n),\,I}-n\o_b\g^{a}\uppsi^{a(n-1)b},\eeq{H2.2}
where repeated indices with the same name are symmetrized with the minimum number of terms and carry unit normalization.

The action~(\ref{H1}) is invariant under the following gauge transformations that involve three 0-form gauge parameters $\uplambda^a$, $\s^a$ and $\upepsilon^{a(n),\,I}$~\cite{Fuentealba:2015jma}:
\bea\d e^a&=&D\uplambda^a-\ve^{abc}\s_be_c+\left(n+\tfrac{1}{2}\right)\bar\upepsilon_{b(n),\,I}\,\g^a\uppsi^{b(n),\,I},\nonumber\\
\d\o^a&=&D\s^a,\label{H3}\\
\d\uppsi^{a(n),\,I}&=&D\upepsilon^{a(n),\,I}-\left(n+\tfrac{1}{2}\right)\s_b\g^b\uppsi^{a(n),\,I}+n\s_b\g^{a}\uppsi^{a(n-1)b},\nonumber\eea{}
where the parameter $\upepsilon^{a(n),\,I}$ is completely symmetric and $\g$-traceless in the frame indices.

Modulo difference in conventions and the flavor index, the action~(\ref{H1}) is formally the same as that of Aragone and Deser~\cite{Aragone:1983sz}. The two, however,
differ in local structure. The local hypersymmetry transformations given in~\cite{Aragone:1983sz} is a subset of~(\ref{H3}) corresponding to the choice:
$\uplambda^a=\s^a=0$ and $\upepsilon^{a(n),\,I}\neq0$, but they agree only on shell.

%
%
%
%
%
%

If we forgo the language of differential forms, the action~(\ref{H1}) takes the form:
\beq S_H=\int d^3x\,e\left[\,\tfrac{2}{\k^2}\,R_{\m\n}{}^{ab}(\o)e_a^\m e_b^\n-\tfrac{1}{2}\,\bar\ps_{\m a(n),\,I}\,\upgamma^{\m\n\r}D_\n\ps_\r{}^{a(n),\,I}\,\right],\eeq{M1}
where $e=\det e_\m^a$, and $\upgamma^{\m\n\r}=\g^{abc}e^\m_a e^\n_b e^\r_c=-e^{-1}\ve^{\m\n\r}$, while
$\ps_\m{}^{a(n),\,I}=2\k^{-1}\,\uppsi_\m{}^{a(n),\,I}$ is the (rescaled) fermion field. The generalized hypersymmetry transformations now read:
\beq \d e_\m^a=\tfrac{1}{4}\!\left(n+\tfrac{1}{2}\right)\k^2\,\bar\e_{b(n),\,I}\,\g^a\ps_\m{}^{b(n),\,I},\qquad \d\ps_\m{}^{a(n),\,I}=D_\m\e^{a(n),\,I},\eeq{hyper12}
where the rescaled transformation parameter is given by: $\e_{a(n),\,I}=2\k^{-1}\,\upepsilon_{a(n),\,I}$\,.

\subsection{Metric-Like Version}\label{sec:M-H}

In this section, we reformulate the frame-like theory of Section~\ref{sec:F-H} in the metric-like language. With this end in view, we switch to the second-order formulation
by integrating out the spin connection $\o_\m{}^{ab}$. In the first-order formulation the equations of motion (EoM) for the spin connection can be solved to obtain a connection
with torsion. This result can then be substituted in the action~(\ref{M1}) to derive the physically equivalent second-order form of the theory with torsion-free connection
and explicit 4-fermion contact terms, exactly as in supergravity~\cite{Freedman:2012zz}. The second-order action for generalized hypergravity looks:
\beq  S_H=\int d^3x\sqrt{-g}\left[\,\tfrac{2}{\k^2}\,R(g)-\tfrac{1}{2}\,\bar\ps_{\m a(n),\,I}\,\upgamma^{\m\n\r}\nabla_\n\ps_\r{}^{a(n),\,I}\,\right]+\text{4-fermion terms},\eeq{M2}
where we have used $D_{[\m}\ps_{\n]}{}^{a(n),\,I}=\nabla_{[\m}\ps_{\n]}{}^{a(n),\,I}$ for torsion-free connection,
and $\upgamma_{\m}\equiv\g_a e^a_\m$. The generalized hypersymmetry transformations are encoded in Eqs.~(\ref{hyper12}).
In particular, note that the transformation rule of the metric tensor, $g_{\m\n}=e_\m^a e_{\n a}$, is given by:
\beq \d g_{\m\n}=\tfrac{1}{2}\!\left(n+\tfrac{1}{2}\right)\k^2\,\bar\e_{a(n),\,I}\,\upgamma_{(\m}\ps_{\n)}{}^{a(n),\,I}.\eeq{M10}

Now we will expand the second-order theory around Minkowski background:
\beq g_{\m\n}=\h_{\m\n}+\k h_{\m\n},\qquad e^a_\m=\bar{e}^{\,a}_\m+\tfrac{1}{2}\k h_\m{}^a+\mathcal O(h^2),\eeq{exp1}
where $\bar{e}^{\,a}_\m$ is the flat-space dreibein: $\bar{e}^{\,a}_\m\bar{e}_{\n a}=\h_{\m\n}$. The fiber indices of the fluctuations, $h_{\m a}$ and $\ps_\m{}^{a(n),\,I}$,
are converted into world indices with the help of the flat-space dreibein $\bar{e}^{\,a}_\m$ and its inverse $\bar{e}^{\,\m}_a$. The metric-like fluctuations include the graviton:
\beq h_{\m\n}\equiv h_{(\m}{}^a\bar{e}_{\n)a}\,,\eeq{exp1.99}
which is a rank-2 symmetric tensor, and the higher-spin Majorana gauge fermion:
\beq \ps_\m{}^{\a(n),\,I}\equiv\ps_\m{}^{a_1\cdots a_n,\,I}\,\bar e^{\,\a_1}_{a_1}\cdots\bar e^{\,\a_n}_{a_n}\,,\eeq{exp2}
which is symmetric and $\g$-traceless in the $\a$-indices. Explicitly, the latter condition reads:
\beq \g_{\b}\ps_\m{}^{\b\a(n-1),\,I}=0,\qquad \g_\b\equiv\g_a\bar{e}^{\,a}_\b\,.\eeq{exp2.5}
Note that the covariant 1-curl of the fermion appearing in the action~(\ref{M2}) is given by:
\beq \nabla_{[\m}\ps_{\n]}{}^{a(n),\,I}=\de_{[\m}\ps_{\n]}{}^{a(n),\,I}+n\o_{[\m}{}^{ab}\ps_{\n]b}{}^{a(n-1),\,I}
+\tfrac{1}{4}\g_{bc}\,\o_{[\m}{}^{bc}\ps_{\n]}{}^{a(n),\,I},\eeq{exp7}
where $\o_\m{}^{ab}=\o_\m{}^{ab}(e)$ is the torsion-free spin connection, whose expansion yields:
\beq \o_{\m a b}=-\tfrac{1}{2}\k\left(\de_a h_{\m b}-\de_b h_{\m a}\right)+\mathcal O(h^2).\eeq{exp8}
The following expressions, with the notation $\displaystyle{\not{\!h\!\,}}_\m=\g^a h_{\m a}=\g^\n h_{\m\n}$, are also necessary:
\bea &\sqrt{-g}=1+\tfrac{1}{2}\k h'+\mathcal O(h^2),\qquad \upgamma^\m=\g^\m+\tfrac{1}{2}\k\displaystyle{\not{\!h\!\,}}_\m+\mathcal O(h^2),&\label{exp2.9}\\
\vspace{10pt}&\upgamma^{\m\n\r}=\g^{\m\n\r}+\tfrac{1}{2}\k\left(\displaystyle{\not{\!h\!\,}}^{[\m}\g^{\n}\g^{\r]}
+\g^{[\m}\displaystyle{\not{\!h\!\,}}^\n\g^{\r]}
+\g^{[\m}\g^\n\displaystyle{\not{\!h\!\,}}^{\r]}\right)+\mathcal O(h^2).&\eea{exp3}

Upon using the expressions~(\ref{exp1})--(\ref{exp3}) in the second-order action~(\ref{M2}), one finds that the desired metric-like theory takes the following form:
\beq S_H=\int d^3x\left[\,\mathcal L_{\text{free}}+\mathcal L_{\text{cubic}}+\mathcal L_{\text{higher-order}}\,\right].\eeq{exp4}
The free part of the Lagrangian $\mathcal L_{\text{free}}$ reads:
\beq \mathcal L_{\text{free}}=\tfrac{1}{2}h_{\m\n}\mathcal G^{\m\n}-\tf{1}{2}\bar{\ps}_{\m\a(n),\,I}\,\mathcal{R}^{\m\a(n),\,I},\eeq{exp5}
where $\mathcal G^{\m\n}$ and $\mathcal{R}^{\m\a(n),\,I}$ are ``the left hand sides'' of the free EoMs:
\begin{equation}\label{exp5.001}
\begin{split}
&\mathcal{G}_{\m\n}\,\equiv\,\Box h_{\m\n}-2\de_{(\m}\de^\r h_{\n)\r}+\de_\m\de_\n h'-\h_{\m\n}\left(\Box h'-\de^\r\de^\s h_{\r\s}\right),\\
&\mathcal{R}^{\m\a(n),\,I}\,\equiv\,\g^{\m\n\r}\de_\n\ps_\r{}^{\a(n),\,I}.
\end{split}
\end{equation}
The cubic Lagrangian $\mathcal L_{\text{cubic}}$ consists of two different kind of vertices: $hhh$-type gravitational self coupling and
$h\ps\ps$-type cross coupling:
\beq \mathcal L_{\text{cubic}}=\mathcal L_{hhh}+\mathcal L_{h\ps\ps}\,.\eeq{exp6}
The graviton cubic self coupling is well known; it is given by (see, for example,~\cite{Boulanger:2000rq}):
\bea \mathcal{L}_{hhh}&=&\k(h^{\m\r}h^{\n\s}\de_\m\de_\n h_{\r\s}-h^{\m\n}h^{\r\s}\de_\m\de_\n h_{\r\s}
+ h^{\m\n}\de_\r h_{\m\s}\de^\r h^\s_\n-\tfrac{1}{2}h^{\m\n}\de_\m h_{\r\s}\de_\n h^{\r\s}\nonumber\\
&&-h^{\m\n}\de^\r h_{\m\n}\de_\r h'-\tfrac{1}{2}h^{\m\r}h^\n_\r\de_\m\de_\n h'
-\tfrac{1}{4}h'\de^\m h^{\n\r}\de_\m h_{\n\r}+h^{\m\n}\de^\r h_{\r\m}\de_\n h'\nonumber\\
&&-h^{\m\n}\de_\r h^\r_\m\de_\s h^\s_\n+\tfrac{1}{2}h^{\m\n}h'\de_\m\de_\n h'
+\tfrac{1}{2}h'\de_\m h^{\m\r}\de_\n h^\n_\r+\tfrac{1}{4}h'\de^\m h'\de_\m h').\eea{self03}
The cubic cross couplings, on the other hand, derive from the fermion-bilinear term in~(\ref{M2}).
As we see in Appendix~\ref{sec:curvatures}, the 1-curl of the fermion, $\de_{[\m}\ps_{\n]\a(n),\,I}$,
itself is proportional to the free EoMs. Then the couplings arising from the first term  in the covariant
1-curl~(\ref{exp7}) are trivial in that they can be field redefined away. The nontrivial cubic couplings
therefore come only from the spin-connection terms in the covariant 1-curl~(\ref{exp7}). They amount to:
\beq \mathcal L_{h\ps\ps}=-\tfrac{1}{4}\k\,\bar{\ps}_{\m\a(n),\,I}\,\g^{\m\n\r}\left(n\h^{\s\l|\a\b}
+\tfrac{1}{4}\g^{\s\l}\h^{\a\b}\right)\ps_{\n\b}{}^{\a(n-1),\,I}\,\mathfrak{h}_{\s\l\Vert\,\r}\,,\eeq{exp9}
where $\mathfrak{h}_{\s\l\Vert\,\r}\equiv\de_\s h_{\l\r}-\de_\l h_{\s\r}$ is the graviton 1-curl, and
$\h^{\s\l|\a\b}\equiv\tfrac{1}{2}\left(\h^{\s\a}\h^{\l\b}-\h^{\s\b}\h^{\l\a}\right)$. On account of the
identity~(\ref{etadefined}) the tensor structure inside the parentheses in Eq.~(\ref{exp9}) can then be
simplified, thanks to the $\g$-trace condition~(\ref{exp2.5}). The final result is:
\beq \mathcal L_{h\ps\ps}=-\tfrac{1}{8}\!\left(n+\tfrac{1}{2}\right)\k\,
\bar{\ps}_{\m\a(n),\,I}\,\g^{\m\n\r}\g^{\s\l}\,\ps_\n{}^{\a(n),\,I}\,\mathfrak{h}_{\s\l\Vert\,\r}\,.\eeq{exp10}

The higher-order couplings $\mathcal L_{\text{higher-order}}$ in the action~(\ref{exp4}) do not stop at any finite order, but they are unique and can also be worked out in principle.
This article, however, does not require any explicit knowledge of these terms. It is important to note that given the cubic couplings, there exists at least one fully
nonlinear consistent theory, namely the generalized hypergravity that we are considering in this section.

Finally, starting from Eq.~(\ref{hyper12}) we would like to write down the generalized hypersymmetry transformations
of the metric-like fluctuations defined in Eqs.~(\ref{exp1}) and~(\ref{exp2}).
It is follows that the graviton transforms as:
\beq \d h_{\m\n}=\tfrac{1}{2}\!\left(n+\tfrac{1}{2}\right)\k\,\bar\e_{\a(n),\,I}\,\g_{(\m}\ps_{\n)}{}^{\a(n),\,I}
+\mathcal O(\k^2),\eeq{exp15}
while the fermion transformation rule is:
\beq \d\ps_\m{}^{\a(n),\,I}=\de_\m\e^{\a(n),\,I}-\tfrac{1}{4}\k\,\mathfrak{h}_{\r\s\Vert\,\m}\left[\left(n+\tfrac{1}{2}\right)
\g^{\r\s}\e^{\a(n),\,I}-n\g_\b{}^{\r\s}\g^\a\e^{\a(n-1)\b,\,I}\right]+\mathcal O(\k^2).\eeq{exp16}

This ends our brief exposition of generalized hypergravity. In the above discussion, when the flavor index $I$ is removed, the theory reduces to $\mathcal N=1$ supergravity and
the spin-$5/2$ Aragone-Deser hypergravity~\cite{Aragone:1983sz} respectively for the cases of $n=0$ and $n=1$.

\section{BRST Deformation Scheme}\label{sec:BRST}

In this section we explain the BRST deformation scheme$-$our main analytical tool to study the uniqueness of hypergravity. What follows is an almost verbatim repetition
of the same discussions appearing in~\cite{Henneaux:2012wg,Henneaux:2013gba}. As pointed out in~\cite{Barnich:1993vg,Henneaux:1997bm}, it is possible reformulate the
classical problem of introducing consistent interactions in a gauge theory in terms of the BRST differential and the BRST cohomology. The advantage of this cohomological
approach is that it systematizes the search for all possible consistent interactions. It also relates the obstructions to deforming a gauge-invariant action to precise
cohomological classes of the BRST differential.

\subsubsection*{Fields and Antifields}

Let there be an irreducible gauge theory of a set of fields $\{\phi^i\}$, with $m$ gauge symmetries: $\d_{\ve}\phi^i=R^i_\a\ve^\a, \a=1,2,...,m$. Then one introduces
a ghost field $\mathcal C^\a$ corresponding to each gauge parameter $\ve^\a$; they have the same algebraic symmetries but opposite Grassmann parity ($\epsilon$). The original
fields and ghosts are collectively called fields, denoted by $\Phi^A$. One further introduces, for each field and ghost, an antifield $\Phi^*_A$ that has the same algebraic
symmetries (in the multi-index $A$) but opposite Grassmann parity.

\subsubsection*{Gradings}

Two gradings are introduced in the algebra generated by the fields and antifields: the pure ghost number ($pgh$) and the antighost number ($agh$). The former is non-zero only
for the ghost fields. For irreducible gauge theories, in particular, one has: $pgh(\mathcal C^\a)=1$, while $pgh(\phi^i)=0$ for any original field. On the other hand, the
antighost number is non-zero only for the antifields $\Phi^*_A$, i.e., $agh(\Phi^*_A)=pgh(\Phi^A)+1,~agh(\Phi^A)=0=pgh(\Phi^*_A)$. Another grading is the ghost number ($gh$),
defined as $gh=pgh-agh$.

\subsubsection*{Antibracket}

On the space of fields and antifields, one then defines an odd symplectic structure:
\beq (X,Y)\equiv\frac{\d^RX}{\d\Phi^A}\frac{\d^LY}{\d\Phi^*_A}-\frac{\d^RX}{\d\Phi^*_A}\frac{\d^LY}{\d\Phi^A}\,.
\eeq{antibracket}
This is called the antibracket that satisfies the graded Jacobi identity. It follows that $\left(\Phi^A,\Phi^*_B\right)=\d^A_B$,
which is real. But a field and its antifield have opposite Grassmann parity. Therefore, if $\Phi^A$ is
purely real (imaginary), $\Phi^*_B$ will be purely imaginary (real).

\subsubsection*{Master Action}

Let $S^{(0)}[\phi^i]$ be the gauge-invariant action in terms of the original fields. One extends it to the master action, $S[\Phi^A,\Phi^*_B]$, that includes terms involving
ghosts and antifields:
\beq S[\Phi^A,\Phi^*_B]=S^{(0)}[\phi^i]+\phi^*_iR^i_\a\mathcal C^\a+\dots\,.\eeq{S_0}
By virtue of the Noether identities and the higher-order gauge-structure equations, it satisfies the classical master equation:
\beq (S,S)=0.\eeq{master}
The master action $S$ incorporates compactly all the consistency conditions pertaining to the gauge transformations through the master equation~(\ref{master}).

\subsubsection*{BRST Differential}

The master action is also the generator of the BRST differential $\mathfrak s$, defined as:
\beq \mathfrak{s}X\equiv(S,X).\eeq{brst1}
It follows, as a simple consequence of the master equation, that $S$ is BRST-closed. From the properties of the antibracket, it is easy to show that $\mathfrak s$
is nilpotent,
\beq \mathfrak s^2=0.\eeq{brst2}
The master action $S$, therefore, belongs to the cohomology of $\mathfrak s$ in the space of local functionals of the fields, antifields, and their finite number
of derivatives.

\subsubsection*{Deformed Master Action}

The existence of the master action $S$ as a solution of the master equation is completely equivalent to the gauge invariance of the original action $S^{(0)}[\phi^i]$.
Therefore, it is possible to reformulate the problem of introducing consistent interactions in a gauge theory as that of deforming the solution $S$ of the master equation.

Let $S$ be the solution of the deformed master equation: $(S,S)=0$. This must be a deformation of the solution $S_0$ of the master equation of the free gauge theory:
\beq S=S_0+gS_1+g^2S_2+\mathcal O(g^3),\eeq{brst5}
in some deformation parameter $g$. The master equation for $S$ splits, up to $\mathcal O(g^2)$,
into \bea (S_0,S_0)&=&0,\label{brst6.1}\\(S_0,S_1)&=&0,\label{brst6.2}\\(S_1,S_1)&=&-2(S_0,S_2).\eea{brst6.3}
Eq.~(\ref{brst6.1}) simply reflects the gauge invariance of the free theory; it also means that $S_0$ is the generator of the BRST differential $s$ of the \emph{free} theory.
Then, Eq.~(\ref{brst6.2}) translates to
\beq s S_1=0,\eeq{brst1st} i.e., the first-order deformation of the master action, $S_1$, is BRST-closed.

\subsubsection*{First-Order Deformations}

Let the first-order local deformations be given by
$S_1=\int a$, where $a$ is a top-form of ghost number 0. Then Eq.~(\ref{brst1st}) gives rise to the cocycle condition: \beq sa\doteq0.
\eeq{cocycle0} Nontrivial deformations therefore belong to $H^0(s|d)$$-$the cohomology of the free
BRST differential $s$, modulo $d$-exact terms, at ghost number 0. One can write~\cite{Boulanger:2000rq,2-s-s,BBH}:
\beq a=a_0+a_1+a_2, \qquad agh(a_i)=i=pgh(a_i),\eeq{brst7}
i.e., the antighost-number expansion of the local form $a$ stops at $agh=2$.
The result is actually more general, and holds for higher-order deformations as well~\cite{Boulanger:2000rq,BBH}.

\subsubsection*{Consistency Cascade}

In terms of the aforementioned gradings, the free BRST differential $s$ splits into the Koszul-Tate differential, $\D$, and the longitudinal derivative along the gauge
orbits, $\G$, as \beq s=\D+\G. \eeq{brst3}
The operator $\D$ implements the equations of motion by acting only on the antifields. It decreases the antighost
number by unity, but keeps the pure ghost number unchanged. On the other hand, $\G$ produces the gauge transformations by acting only on the original fields.
It increases the pure ghost number by unity without changing the antighost number. Thus, $gh(\D)=gh(\G)=gh(s)=1$. Note that $\D$ and $\G$ are nilpotent and anticommuting,
\beq \G^2=\D^2=0,\qquad\G\D+\D\G=0.\eeq{brst4}

Given the expansion~(\ref{brst7}) and the splitting~(\ref{brst3}), the cocycle condition~(\ref{cocycle0}) gives rise to
the following cascade of relations consistent first-order deformations must obey:
\bea \G a_2&\doteq&0,\label{cocycle1}\\
\D a_2+\G a_1&\doteq&0,\label{cocycle2}\\
\D a_1+\G a_0&\doteq&0.\eea{cocycle3}
%
The set of conditions~\rf{cocycle1}--\rf{cocycle3} is dubbed
the consistency cascade. Note that one can always choose $a_2$ to be $\G\text{-closed}$, instead of $\G\text{-closed}$ modulo $d$~\cite{BBH}.

\subsubsection*{More on First-Order Deformations}

The various terms in the expansion~(\ref{brst7}) of the first-order deformations have the following significance. The term $a_0$ gives the deformation
of the Lagrangian, while $a_1$ and $a_2$ capture the deformations of the gauge transformations and the gauge algebra respectively~\cite{Barnich:1993vg,Henneaux:1997bm}.
Indeed, a nontrivial $a_2$ implies the deformation of the gauge algebra into a non-Abelian one.
Note that $a_2$ will be trivial iff it can be removed by adding $s$-exact modulo $d$-exact terms: $sm+dn$. Expanding $m$ and $n$
in antighost number and considering the fact that $m$ and $n$ also stop at $agh=2$, one concludes that $a_2$ is trivial iff $a_2=\G m_2 + dn_2$. Thus, the cohomology of
$\G$ modulo $d$ plays an important role. The gauge algebra will be deformed if and only if $a_2$ is a nontrivial element of the cohomology of $\G$ modulo $d$.

On the other hand, if $a_2$ is trivial, the algebra remains Abelian up to first-order deformations. In this case, one can always choose $a_2=0$, and $\G a_1=0$~\cite{BBH}.
A non-trivial $a_1$ then implies that the gauge transformations are deformed while the algebra remains Abelian. Again, $a_1$ is trivial if it is $\D$-exact modulo $d$,
$a_1=\D m_2+dn_1$. In this case, $a_1$ can be removed, so that one can choose $a_0$ to be $\G$-closed modulo $d$: the action is deformed but the gauge transformations
remain undeformed. Lagrangian deformations $a_0$ are (non)trivial iff they are (non)trivial elements in the cohomology $H^0(\D|d)$. Therefore, two vertices are equivalent
iff they differ only by $\D$-exact modulo $d$ terms.

\subsubsection*{Second-Order Deformations}

The second-order consistency condition~\rf{brst6.3} requires that $(S_1,S_1)$ be $s$-exact\footnote{$(S_1,S_1)$ is BRST-closed  automatically;
it follows from the graded Jacobi identity for the antibracket.},
\beq (S_1,S_1)=-2s S_2.\eeq{brst2nd}
Let us consider the following expansions in antighost number:
\beq S_2=\int\left(b_0+b_1+b_2\right),\qquad (S_1,S_1)=-2\int\left(c_0+c_1+c_2\right),\eeq{new1}
where the $b_i$'s incorporate the second-order deformations, and the $c_i$'s are given by:
\bea \int\!c_2&=&-\tfrac{1}{2}\left(\int\!a_2\,,\int\!a_2\right),\label{new2.1}\\
 \int\!c_1&=&-\left(\int\!a_2\,,\int\!a_1\right)-\tfrac{1}{2}\left(\int\!a_1\,,\int\!a_1\right),\label{new2.2}\\
 \int\!c_0&=&-\left(\int\!a_1\,,\int\!a_0\right).\eea{new2.3}

In view of Eqs.~(\ref{brst2nd})--(\ref{new1}) and the splitting~(\ref{brst3}), we obtain the following set of relations that consistent second-order
deformations must fulfill:
\bea c_2&\doteq&\G b_2,\label{new3.1}\\
c_1&\doteq&\D b_2+\G b_1,\label{new3.2}\\
c_0&\doteq&\D b_1+\G b_0.\eea{new3.3}
These conditions determine whether or not, in a local theory, a consistent first-order deformations get obstructed at the second order. Such higher-order obstructions
are controlled by the local BRST cohomology group $H^1(s|d)$~\cite{BBH}.

\section{Uniqueness of Spin-5/2 Hypergravity}\label{sec:Metric}

In this section we take recourse to the metric-like formulation of higher spins, and consider consistent interactions of a spin-2 and a Majorana spin-5/2 gauge fields in
flat space. We employ the BRST deformation scheme outlined in the previous section under the assumptions of locality, parity and Poincar\'e invariance.
Nontrivial interaction terms of the gauge fields are restricted by the requirement that gauge symmetries be preserved modulo possible deformations. We would however
postpone until Section~\ref{sec:CC} the consideration of a (linearized) cosmological term that appears naturally as a tadpole in the first-order Lagrangian
deformation~\cite{Boulanger:2000rq}.

\subsection{The Free Theory}\label{subsec:5half-free}

Our starting point is the free theory, which is a special case of Eqs.~(\ref{exp5})--(\ref{exp5.001}) with $n=1$ and the omission of the flavor index:
\beq \mathcal L_{\text{free}}=\tfrac{1}{2}h_{\m\n}\mathcal G^{\m\n}-\tf{1}{2}\bar{\ps}_{\m\a}\,\mathcal{R}^{\m\a},\eeq{3.1}
where $\mathcal G^{\m\n}$ is the linearized Einstein tensor defined in Eq.~(\ref{exp5.001}), and $\mathcal{R}^{\m\a}\equiv\g^{\m\n\r}\de_\n\ps_\r{}^\a$.
Here, the spin-2 gauge field $h_{\m\n}$ is a symmetric rank-2 tensor, whereas the spin-$5/2$  Majorana gauge fermion $\ps_{\m\a}$
has no symmetry in its tensor indices $\mu$ and $\a$, but is $\g$-traceless w.r.t.~the second index:
\beq \g^\a\ps_{\m\a}=0.\eeq{3.2}
The Lagrangian~(\ref{3.1}) enjoys the following Abelian gauge symmetries:
\beq \d_\l h_{\m\n}=2\de_{(\m}\l_{\n)},\qquad \d_\e\ps_{\m\a}=\de_\m\e_\a,\eeq{gaugeinv2}
where $\l_\m$ is a vector parameter, and $\e_\a$ is a $\g$-traceless Majorana vector-spinor: $\displaystyle{\not{\!\e}}=0$.

Because $\ps_{\m\a}$ is not symmetric in the tensor indices, it is clearly not a Fronsdal tensor-spinor. It is actually a ``dreibein-like fermion''
with covariant indices. However, the two are connected by the simple relation~\cite{Rahman:2017cxk}:
\beq \ps_{\m\n}=\Ps_{\m\n}+2\g_{[\m}\Pss_{\n]}-\tfrac{1}{2}\g_{\m\n}\Ps',\eeq{ad2f}
where $\Ps_{\m\n}=\Ps_{\n\m}$ is a Fronsdal field. It is easy to see that the relation~(\ref{ad2f})$-$compatible with the
constraint~(\ref{3.2})$-$maps the fermionic part of
the Lagrangian~(\ref{3.1}) into the well-known spin-5/2 Fang-Fronsdal Lagrangian (see~\cite{Rahman:2015pzl,Rahman:2017cxk} for recent reviews):
\bea \mathcal L_\text{FF}=-\tfrac{1}{2}\bar{\Ps}_{\m\n}\!\!\not{\!\partial\!}\;\Ps^{\m\n}-\bar{\displaystyle\not{\!\!\Ps}}_{\m}\!\not{\!\partial\!}\,
\displaystyle\not{\!\!\Ps}^{\mu}+\tfrac{1}{4}\bar{\Ps}'\!\not{\!\partial\!}\;\,\Ps'+\tfrac{1}{2}\,\bar{\displaystyle\not{\!\!\Ps}}_\m
\,\partial\!\cdot\!\Ps^\m-\bar{\Ps}'\partial\,\cdot\!\displaystyle\not{\!\!\Ps}.\eea{F-lagr}

The two descriptions are in fact completely equivalent~\cite{Rahman:2017cxk}. It is well known that the Fang-Fronsdal spin-5/2 system~(\ref{F-lagr}),
with its own gauge invariance, has zero degrees of freedom in $D=3$ (see, for example,~\cite{Rahman:2015pzl}). To see that this count remains
the same for our ``dreibein-like field''~\cite{Rahman:2017cxk}, let us first choose the covariant gauge:
\beq \mathcal G^{(1)}_\a\equiv\g^\m\ps_{\m\a}=0,\eeq{3.5}
by making use of the freedom of the gauge parameter $\e_\a$. Then, by virtue of identities~(\ref{3gamma1})--(\ref{2gamma1}),
the Euler-Lagrange equations of motion:
$\mathcal R^{\m\a}=0$, and its $\g$-trace: $\g_\m\mathcal{R}^{\m\a}=0$, yield the Dirac equation plus the divergence constraint:
\beq \ds\,\ps_{\m\a}=0,\qquad \de^\m\ps_{\m\a}=0.\eeq{3.6}
The gauge fixing~(\ref{3.5}), however, is not complete. The residual gauge freedom can be exhausted by further choosing:
\beq \mathcal G^{(2)}_\a\equiv\ps_{0\a}=0.\eeq{3.7}
The count of local physical degrees of freedom is now immediate. The system of equations~(\ref{3.6}) describes $(3-1)\times2=4$ dynamical variables.
However, each of the gauge choices~(\ref{3.5}) and~(\ref{3.7}) removes 2 degrees of freedom.
Therefore, the total number of physical degrees of freedom is $(4-2-2)=0$, as it should be.

Having at hand the gauge-invariant action of the original fields,
\beq S^{(0)}\left[h_{\m\n}, \ps_{\m\a}\right]=\int d^3x\left(\tfrac{1}{2}h_{\m\n}\mathcal G^{\m\n}-\tf{1}{2}\bar{\ps}_{\m\a}\,\mathcal{R}^{\m\a}\right),
\eeq{free-action}
we will now construct the free master action. With this end in view, we introduce the Grassmann-odd bosonic ghost $C_\m$ for the Grassmann-even bosonic gauge parameter $\l_\m$.
Likewise, corresponding to the Grassmann-odd real Majorana-spinor gauge parameter $\e_\a$, we have the Grassmann-even Majorana-spinor ghost $\xi_\m$, which is of
course $\g$-traceless. Thereby, the set of fields is augmented to
\beq \Phi^A=\{h_{\m\n}, C_\m, \ps_{\m\a}, \xi_\a\}.\eeq{fieldsset5/2}
For each of these fields, we introduce an antifield with the same algebraic symmetries in its indices, but opposite Grassmann parity and opposite phase in the
Majorana condition for spinors (we choose $\b=1$ for fields, and $\b=-1$ for antifields). The set of antifields is
\beq \Phi^*_{A}=\{h^{*\m\n}, C^{*\m}, \bar{\ps}^{*\m\a},\bar{\xi}^{*\a}\}.\eeq{antifieldsset5/2}

In Table 1, we enumerate the various fields and antifields along with the actions of $\G$ and $\D$ on them, spell out their gradings, Grassmann parity $\e$, and
the phase $\b$ in the Majorana condition (for spinors).
\begin{table}[ht]
\caption{Properties of the Fields \& Antifields ($s=5/2$)}
\vspace{6pt}
\centering
\begin{tabular}{c c c c c c c c}
\hline\hline
$Z$ &$\G(Z)$~~~&~~~$\D(Z)$~~~&$pgh(Z)$ &$agh(Z)$ &$gh(Z)$ &$\epsilon(Z)$ &$\beta(Z)$\\ [0.5ex]
\hline
$h_{\m\n}$ & $2\de_{(\m} C_{\n)}$ & 0 & 0 & 0 & 0 & 0 &-\\
$C_\m$ & 0 & 0 & 1 & 0 & 1 & 1 &-\\
$h^{*\m\n}$ & 0 & $\mathcal G^{\m\n}$ & 0 & 1 & $-1$ & 1 &-\\
$C^{*\m}$ & 0 & $-2\de_\n h^{*\m\n}$ & 0 & 2 & $-2$ & 0 &-\\ \hline
$\ps_{\m\a}$ & $\de_\m\xi_\a$ & 0 & 0 & 0 & 0 & 1 &$+1$\\
$\xi_\a$ & 0 & 0 & 1 & 0 & 1 & 0 &$+1$\\
$\bar{\ps}^{*\m\a}$ & 0 & $-\bar{\mathcal R}^{\m\a}$ & 0 & 1 & $-1$ & 0 &$-1$\\
$\bar{\xi}^{*\a}$ & 0 & $\de_\m\bar{\ps}^{*\m\a}$ & 0 & 2 & $-2$ & 1 &$-1$\\
\hline\hline
\end{tabular}
\end{table}
\vspace{6pt}

The free master action $S_0$ is an extension of the original gauge-invariant action~\rf{free-action} by terms involving ghosts and antifields. Explicitly,
\beq S_0=\int d^3x\Big(\tfrac{1}{2}h_{\m\n}\mathcal G^{\m\n}-\tf{1}{2}\bar{\ps}_{\m\a}\mathcal{R}^{\m\a}
+2h^{*\m\n}\de_\m C_\n -\bar{\ps}^{*\m\a}\de_\m\xi_\a\Big).\eeq{freemasteraction5/2}
Notice that the antifields source the gauge variations, with gauge parameters replaced by corresponding ghosts. It is easy to see that $S_0$ satisfies
the master equation: $(S_0,S_0)=0$.

\subsection{First-Order Deformations}\label{subsec:5half-1st}

We are now ready to study the deformations of the master action that are first order in the infinitesimal parameter $g$.
Apart from locality and Poincar\'e invariance, we will assume that the deformed theory preserves parity, which is a symmetry
of the free theory~(\ref{freemasteraction5/2}). First, we will consider deformations that
correspond to non-Abelian vertices in the theory. Then we will study Abelian vertices, and show that in $D=3$ such vertices cease to exist.

\subsubsection{Non-Abelian Vertices}\label{subsec:nonAbelian}

Non-Abelian vertices deform the gauge algebra, and correspond to deformations of the master action with $a_2$ being a nontrivial
element in $H(\G|d)$. Note that $a_2$ is a Grassmann-even, Hermitian and parity-invariant
(by assumption) Lorentz scalar with a vanishing ghost number: $gh(a_2)=0$. By invoking the same logic as that appearing
in Ref.~\cite{Boulanger:2000rq}, it is easy to see that an $a_2$ will consist of a single antighost and two ghost fields. Non-Abelian
first-order deformations will therefore give rise to cubic vertices in the theory.

Two $a_2$'s are equivalent if and only is they differ by $\G$-exact terms modulo $d$. Then, without loss of generality, we can choose the antighost appearing in $a_2$ to be
undifferentiated. We see from Appendix~\ref{sec:cohomology} that any derivative acting on the bosonic ghost field $C_\m$ can be realized as a
1-curl $\mathfrak C_{\m\n}$ up to irrelevant $\G$-exact terms. On the other hand, derivatives of the fermionic ghost $\xi_\a$ are always trivial
in the cohomology of $\G$. The possible types of non-Abelian vertices are twofold:
graviton self coupling and cross coupling.
\vspace{5pt}
\newline
\noindent{\bf Graviton Self Coupling:} Because any derivative of the ghost-curl $\mathfrak C_{\m\n}$ is $\G$-exact, nontrivial $a_2$'s for graviton
self coupling may contain only up to 2 derivatives. Parity and Poincar\'e invariance, however, leave us with the unique
possibility~\cite{Boulanger:2000rq}:
\beq a_2\,=\,\l C_\m^*C_\n\de^\m C^\n,\eeq{self1}
where $\l$ is a coupling constant. Upon using the second equation~(\ref{cocycle2})
of the consistency cascade, after a straightforward calculation one arrives at~\cite{Boulanger:2000rq}:
\beq a_1\,=\,\l h^*_{\m\n}C_\r\left(\de^\r h^{\m\n}-\de^\m h^{\n\r}-\de^\n h^{\m\r}\right)+\tilde{a}_1,\eeq{self2}
where the ambiguity term $\tilde{a}_1$ is a nontrivial element in $H(\G)$ with $agh(\tilde{a}_1)=1$
and $gh(\tilde{a}_1)=0$. As shown in Appendix~\ref{sec:cohomology}, one must have $\tilde{a}_1=\D\text{-exact}$.
Given the cocycle condition~(\ref{cocycle3}), such an $\tilde{a}_1$ has no bearing whatsoever in the vertex $a_0$,
and therefore can be set to zero without any loss of generality. This is in contrast with the case of
$D>3$~\cite{Boulanger:2000rq}, where there exist nontrivial ambiguity terms that may give rise to higher-derivative
cubic couplings unless restriction on the number of derivatives is assumed.

Then, from the third equation~(\ref{cocycle3}) of the consistency cascade one arrives at~\cite{Boulanger:2000rq}:
\beq a_0\,=\,\l\left(\k^{-1}\mathcal{L}_{hhh}\right),\eeq{self3}
where $\mathcal{L}_{hhh}$ has been given in Eq.~(\ref{self03}). Therefore, up to an overall coupling constant, the
graviton cubic self coupling is the same that appears in General Relativity. Moreover, this is the off-shell form of
the only parity-preserving vertex that one would expect in $D=3$ from the classification of cubic
vertices of bosonic gauge fields~\cite{Mkrtchyan:2017ixk,Kessel:2018ugi}.
\vspace{5pt}
\newline
\noindent{\bf Cross Coupling:} Nontrivial $a_2$'s for cross coupling, on the other hand, cannot contain more than 1 derivative.
The most general gauge-algebra deformation reads:
\beq a_2\,=\,\b_0\,\bar\xi^\a\g^\m\xi^*_\a C_\m+\b_1\,\bar\xi_\a\g^\m\xi^\a C^*_\m+\b_2\,\bar\xi_\a\g^{\m\n}\xi^{*\a}\mathfrak{C}_{\m\n}.\eeq{gad}
At this point, it is important to take note of the $\g$-matrix identities~(\ref{etadefined}) and~(\ref{3gamma1}). In $D=3$, in view of the
$\g$-tracelessness of the fermionic (anti)ghost, one finds that bilinears containing 0 or 1 $\g$-matrix are equivalent those containing
respectively 2 or 3 $\g$-matrices, e.g., $\bar\xi_{[\m}\,\xi^*_{\n]}=\tfrac{1}{2}\,\bar\xi_\a\g_{\m\n}\,\xi^{*\a}$, and
$\bar\xi_\a\g^\m\xi^\a=-\bar\xi_\a\g^{\m\a\b}\xi_\b$ etc. It is therefore clear that the $a_2$ given in Eq.~(\ref{gad}) did not miss any
linearly independent terms.

Now, taking the $\D$-variation of Eq.~(\ref{gad}), one arrives at the following form:
\beq \D a_2\,\doteq\,\tfrac{1}{2}\b_0\,\bar\xi_\a\g^\m\ps^{*\n\a}\mathfrak{C}_{\m\n}+\G\text{-exact}.\eeq{toa0.5}
The first term proportional to $\b_0$ is nontrivial in the cohomology of $\G$ modulo $d$.
In order for the second equation~(\ref{cocycle2}) of the consistency cascade to be fulfilled, we must set:
\beq \b_0=0.\eeq{g0value}
This choice reduces Eq.~(\ref{toa0.5}) to the desired form:
\beq \D a_2\,\doteq\, 4\b_1\,\bar\xi_\a\g^\m\de^\n\xi^\a h^*_{\m\n}-\b_2\left(\bar\xi_\a\g^{\m\n}\ps^{*\r\a}\de_\r\mathfrak{C}_{\m\n}
+\de_\r\bar\xi_\a\g^{\m\n}\ps^{*\r\a}\mathfrak{C}_{\m\n}\right),\eeq{toa1}
where the right-hand side is manifestly $\G$-exact. Thus, we arrive at:
\beq a_1\,=\,-4\b_1\,\bar\xi_\a\g^\m\ps^{\n\a} h^*_{\m\n}+\b_2\left(\bar\xi_\a\g^{\m\n}\ps^{*\r\a}\mathfrak{h}_{\m\n\Vert\r}+
\bar\ps_{\r\a}\g^{\m\n}\ps^{*\r\a}\mathfrak{C}_{\m\n}\right),\eeq{gtd}
where again we have dropped the ambiguity term $\tilde{a}_1\in H(\G)$. The reasoning behind this is already given
in the study of graviton self coupling.

Taking again a $\D$-variation and using $\D\ps^{*\r\a}=\mathcal{R}^{\r\a}=-\ve^{\r\s\l}\de_\s\ps_\l{}^\a$, we get:
\beq \D a_1\,=\,4\b_1\,\bar\xi_\a\g^\m\ps^{\n\a}\mathcal{G}_{\m\n}-\b_2\,\ve^{\r\s\l}\left(\bar\xi_\a\g^{\m\n}\de_\s\ps_\l{}^\a
\mathfrak{h}_{\m\n\Vert\r}-\bar\ps_{\r\a}\g^{\m\n}\de_\s\ps_\l{}^\a\mathfrak{C}_{\m\n}\right).\eeq{toa0}
By virtue of the relation: $\mathcal G^{\t\l}=\tfrac{1}{4}\ve^{\t\m\n}\ve^{\l\r\s}R_{\m\n\r\s}$,
the first term in Eq.~(\ref{toa0}) can be written in terms of the linearized Riemann tensor. On the other hand, for Majorana spinors one can write:
$\ve^{\r\s\l}\bar\ps_{\r\a}\g^{\m\n}\de_\s\ps_\l{}^\a=\tfrac{1}{2}\ve^{\r\s\l}\de_\s\left(\bar\ps_{\r\a}\g^{\m\n}\ps_\l{}^\a\right)$,
which simplifies the last term. Up to total derivatives therefore we can write:
\beq
\D a_1\,\doteq\,-\left(\b_1+\tfrac{1}{2}\b_2\right)\ve^{\r\s\l}\bar\xi_\a\g^{\m\n}\ps_\l{}^\a R_{\m\n\r\s}\,+\,\G\text{-exact}.
\eeq{toa2}
The first term in the above equation is nontrivial in the cohomology of $\G$ modulo $d$, and so its coefficient must be set to zero in order that Eq.~(\ref{cocycle3}) be fulfilled:
\beq \tfrac{1}{2}\b_2\,=\,-\b_1\,\equiv\,\b\,.\eeq{toa3}
Then, the right hand side of Eq.~(\ref{toa2}) is left with the desired $\G$-exact piece:
\beq \D a_1\doteq-\b\,\ve^{\r\s\l}\left(2\de_\r\bar\xi_\a\g^{\m\n}\ps_\l{}^\a
\mathfrak{h}_{\m\n\Vert\s}+\bar\ps_{\r\a}\g^{\m\n}\ps_\l{}^\a\de_\s\mathfrak{C}_{\m\n}\right),\eeq{toa4}
so that in accordance with Eq.~(\ref{cocycle3}) we finally obtain the cubic cross coupling:
\beq a_0\,=\,\b\,\bar{\ps}_{\m\a}\,\g^{\m\n\r}\g^{\s\l}\,\ps_\n{}^\a\,\mathfrak{h}_{\s\l\Vert\,\r}\,.\eeq{toa5}
Note that this is a parity-preserving 1-derivative vertex, and as such it qualifies as the minimal coupling of
a massless spin-5/2 field to a spin-2 gauge field.

\subsubsection{Absence of Abelian Vertices}\label{subsec:abelian}

We will now show that there are no Abelian verices\footnote{We are interested in vertices that are at least cubic in the fields.
In other words, we are not considering tadpole terms, which however will be taken into account in Section~\ref{sec:CC}.} for the
system under consideration. Abelian vertices are
those that do not deform the gauge algebra, and correspond to trivial $a_2$'s. Then, it is always possible to choose
an $a_1$ that is $\G\text{-closed}$~\cite{BBH}:
\beq \G a_1=0,\eeq{noa2strict}
and relates to the Abelian vertex $a_0$ through the cocycle condition~(\ref{cocycle3}). However, it follows from
Appendix~\ref{sec:cohomology} that $a_1=\D\text{-exact}$.
In view of the cocycle condition~(\ref{cocycle3}), a $\D$-exact $a_1$ does not affect the vertex $a_0$. Therefore,
without any loss of generality, $a_1$ can be set to zero. The cocycle condition~(\ref{cocycle3}) then reduces to
the following form:
\beq \G a_0\doteq0,\eeq{noa2strict3}
from which it follows that the vertex cannot be nontrivial, i.e., $a_0$ must be of the form:
\beq a_0\doteq\D\text{-exact}.\eeq{noa2strict3.1}

To see this, let us note that it is always possible to rewrite a vertex as one of the fields
$\{h_{\m\n}, \bar\ps_{\m\a}\}$ contracted respectively with one of the currents $\{T^{\m\n}, \Th^{\m\a}\}$:
\beq a_0=\left\{
           \begin{array}{ll}
             h_{\m\n}T^{\m\n}, &  \text{gravitons appear}\\
             \bar\ps_{\m\a}\Th^{\m\a}, &  \text{otherwise.}
           \end{array}
         \right.\eeq{noa2strict4}
Then, inserting this form of $a_0$ into Eq.~(\ref{noa2strict3}) we respectively arrive at:
\beq h_{\m\n}\,\G T^{\m\n}-2C_\n\de_\m T^{\m\n}\doteq0,\qquad
\bar\ps_{\m\a}\G\Th^{\m\a}+\x_\a\de_\m\Th^{\m\a}\doteq0.\eeq{noa2strict5}
Note that the ghost fields appear in $\G T^{\m\n}$ and $\G\Th^{\m\a}$ only \emph{with} derivatives. A functional derivative of Eq.~(\ref{noa2strict5}) w.r.t.~an \emph{undifferentiated} ghost then yields:
\beq \de_\m T^{\m\n}=0,\qquad \de_\m\Th^{\m\a}=0.\eeq{noa2strict5.1}
In order for the currents to be identically conserved, they must take the form:
\beq T^{\m\n}=\de_\r\de_\s X^{[\m\r][\n\s]},\qquad \Th^{\m\a}=\de_\n Y^{[\m\n]\a}.\eeq{noa2strict5.2}
Here $X$ has the symmetry of the window diagram {\tiny$\young(\null\null,\null\null)$}, i.e., that of the Riemann tensor.
While the antisymmetry of $X^{[\m\r][\n\s]}$ in the indices $(\m,\r)$ and $(\n,\s)$ is required by the identical conservation
of the current, symmetry in the indices $(\r,\s)$ is imposed the derivatives in front, which in turn demands symmetry in
$(\m,\nu)$. On the other hand, $Y^{[\m\n]\a}$ is antisymmetric in the indices $(\m,\n)$ and $\g$-traceless in the index $\a$.
Plugging the required forms~(\ref{noa2strict5.2}) of the currents into Eq.~(\ref{noa2strict4}), and the integrating by parts,
we arrive at:
\beq a_0\doteq\left\{
           \begin{array}{ll}
             \tfrac{1}{4}R_{\m\n\r\s}X^{[\m\n][\r\s]}, &  \text{gravitons appear}\\
             -\de_{[\m}\bar\ps_{\n]\a}Y^{[\m\n]\a}, &  \text{otherwise.}
           \end{array}
         \right.\eeq{noa2strict5.3}
Because the curvatures $\{R_{\m\n\r\s}, \de_{[\m}\ps_{\n]\a}\}$ are $\D$-exact (see Appendix~\ref{sec:curvatures}),
it follows that an Abelian vertex $a_0$ must be $\D$-exact modulo $d$ as claimed\footnote{Because we did not assume
parity invariance of $T^{\m\nu}$, it follows as a byproduct of our analysis that the parity non-invariant 3-derivative
cubic self coupling of gravitons, reported in~\cite{Kessel:2018ugi}, must be non-Abelian.}.

\subsubsection{Summary at First Order}\label{subsec:summary1st}

Let us make a summary of the results obtained in this section. The complete set of first-order deformations of the
master action is given by:
\bea a_2&=&\,\l C_\m^*C_\n\de^\m C^\n+\b\left(2\bar\xi_\a\g^{\m\n}\xi^{*\a}\mathfrak{C}_{\m\n}-\bar\xi_\a\g^\m\xi^\a
C^*_\m\right),\label{2nd1}\\
a_1&=&\l h^*_{\m\n}C_\r\left(\de^\r h^{\m\n}-\de^\m h^{\n\r}-\de^\n h^{\m\r}\right)\nonumber\\
&&+2\b\left(\bar\ps_{\r\a}\g^{\m\n}\ps^{*\r\a}\mathfrak{C}_{\m\n}+2\bar\xi_\a\g^\m\ps^{\n\a} h^*_{\m\n}+\bar\xi_\a\g^{\m\n}\ps^{*\r\a}\mathfrak{h}_{\m\n\Vert\r}\right),\label{2nd1.1}\\
a_0&=&\l\left(\k^{-1}\mathcal{L}_{hhh}\right)+\b\,\bar{\ps}_{\m\a}\,\g^{\m\n\r}\g^{\s\l}\,\ps_\n{}^\a\,
\mathfrak{h}_{\s\l\Vert\,\r}\,,\eea{2nd1.2}
where $\l$ and $\b$ are two yet-arbitrary coupling constants, and $\mathcal{L}_{hhh}$ is given by Eq.~(\ref{self03}).

\subsection{Consistency at Second Order}\label{subsec:1-def}

We will now study the consistency requirements at second-order in the deformation parameter $g$.
First, we would like to compute the quantity $c_2$ given through Eq.~(\ref{new2.1}) by the antibracket of $\int\!a_2$
with itself. Because $\int\!a_2$ is Grassmann even, we can write:
\beq \int c_2=-\left(\frac{\d^R}{\d\Phi^A}\int a_2\right)\left(\frac{\d^L}{\d\Phi^*_A}\int a_2\right).\eeq{2nd2}
Schematically, $c_2$ can be written as a sum of terms proportional to $\a^2$, $\a\b$ and $\b^2$:
\beq c_2=\l^2c_2^{(\l\l)}+\l\b\,c_2^{(\l\b)}+\b^2c_2^{(\b\b)},\eeq{2nd3}
where $c_2^{(\l\l)}$, $c_2^{(\l\b)}$ and $c_2^{(\b\b)}$ are Lorentz scalars having the same gradings as $c_2$.
While computing these quantities, it is important thing to keep in mind that a functional derivative w.r.t.~the
fermionic (anti)ghost must be $\g$-traceless. A straightforward calculation yields:
\bea &c_2^{(\l\l)}=\left[\de^\m\!\left(C^*_\m C_\n\right)+C^*_\m\de^\m C_\n\right]C_\r\de^\n C^\r,&\label{2nd4.1}\\
&c_2^{(\l\b)}=-\left[\de^\m\!\left(C^*_\m C_\n\right)+C^*_\m\de^\m C_\n\right]\bar\xi_\a\g^\n\xi^\a
+4\de_\m\!\left(\bar\xi_\a\g^{\m\n}\xi^{*\a}\right)C_\r\de_\n C^\r,&\label{2nd4.2}\\
&c_2^{(\b\b)}=-4\left(\bar\xi^*_\a\g^{\m\n}\mathfrak{C}_{\m\n}+\bar\xi_\a\g^\m C^*_\m\right)_{\g\text{-}t} \left(\g^{\r\s}\xi^\a\mathfrak{C}_{\r\s}\right)_{\g\text{-}t},&\eea{2nd4.3}
where the subscript ``$\g\text{-}t$'' stands for the $\g$-traceless part of the term under consideration.

The second-order deformations ought to fulfill the consistency conditions~(\ref{new3.1})--(\ref{new3.3}).
Then, it is necessary that $c_2$, given in Eqs.~(\ref{2nd3})--(\ref{2nd4.3}), be trivial in the
cohomology $H^1(\G|d)$. In other words, we would like to find a quantity $b_2$, with $agh(b_2)=2$
and $gh(b_2)=0$, such that $c_2\doteq\G b_2$. The requirement that such a $b_2$ exist poses nontrivial constrains
on the cubic couplings $\l$ and $\b$, as we will see.

Now, the consistency of General Relativity as a gauge theory ensures that $c_2^{(\l\l)}$ itself belongs to $H^1(\G|d)$.
It is not difficult to reconfirm this. Indeed, one can rewrite Eq.~(\ref{2nd4.1}) in terms of an undifferentiated antighost, as:
\bea
c_2^{(\l\l)}&\doteq& C^*_\m \left(-C_\n\de^\m C_\r\de^\n C^\r-C_\n C_\r\de^\m \de^\n C^\r+\de^\m C_\n C_\r\de^\n C^\r\right)\nonumber\\
&=&C^*_\m \left[-2C_\n C_\r\de^\n\de^{(\m}C^{\r)}+2\de^\m C^\n C^\r\de_{(\n}C_{\r)}\right],
\eea{2nd5}
where in the last line we have used that fact that $C_\m$ is Grassmann odd. This gives:
\beq c_2^{(\l\l)}\doteq \G b_2^{(\l\l)},\qquad b_2^{(\l\l)}\equiv
C^*_\m \left(\de^\m C^\n C^\r h_{\n\r}-C_\n C_\r\de^\n h^{\m\r}\right).\eeq{2nd6}

Next we consider Eq.~(\ref{2nd4.2}), and rewrite $c_2^{(\l\b)}$ in terms of undifferentiated antighosts,
modulo total derivatives. Then, on account of Eqs.~(\ref{Gammaaction2}) and~(\ref{g1.9}), we arrive at:
\beq c_2^{(\l\b)}\doteq \bar\xi^*_\a\g^{\m\n}\xi^\a\mathfrak{C}_{\m\r}\mathfrak{C}_\n{}^\r
-\tfrac{1}{2}C^*_\m\bar\xi_\a\g_\n\xi^\a\mathfrak{C}^{\m\n}+\G\text{-exact}.\eeq{2nd7}
The first two terms appearing on the right side of in Eq.~(\ref{2nd7}) are nontrivial elements in $H^1(\G|d)$.
Such terms must be eliminated in order for the theory to be consistent beyond first order in the deformation
parameter. The only way this may happen, if at all, is through the mutual cancellation of similar terms arising
possibly from $c_2^{(\b\b)}$.

To see if the offending terms in $c_2^{(\l\b)}$ can be eliminated, we need to compute $c_2^{(\b\b)}$ from
Eq.~(\ref{2nd4.3}). The technical steps of the explicit computation are relegated to Appendix~\ref{sec:details}.
The result turns out to be exactly of the form we have hoped for. To be explicit,
\beq c_2^{(\b\b)}=\tfrac{16}{3}\left(\bar\xi^*_\a\g^{\m\n}\xi^\a\mathfrak{C}_{\m\r}\mathfrak{C}_\n{}^\r
-\tfrac{1}{2}C^*_\m\bar\xi_\a\g_\n\xi^\a\mathfrak{C}^{\m\n}\right)+\G\text{-exact}.\eeq{2nd9}

In view of the expressions~(\ref{2nd6})--(\ref{2nd9}), we conclude from Eq.~(\ref{2nd3}) that the resulting $c_2$
will be a trivial element in $H^1(\G|d)$ if and only if the couplings $\l$ and $\b$ satisfy:
\beq \b\left(\l+\tfrac{16}{3}\b\right)=0.\eeq{2nd10}
Therefore, the absence of obstructions at $\mathcal O(g^2)$ necessarily requires that the number of independent coupling
constants appearing at $\mathcal O(g)$ reduces to a single one. The quadratic constraint~(\ref{2nd10}) gives rise to the
following two branches of solutions:
\bea \b=\left\{
  \begin{array}{ll}
    -\tfrac{3}{16}\l, & \text{fermion coupled to gravity}\\
    0, & \text{free fermion plus gravity.}
  \end{array}
\right.\eea{2nd11}
We would be interested in the first branch, which corresponds to a non-vanishing $\b$ and therefore to
a nontrivial coupling to gravity of the spin-5/2 fermion at $\mathcal O(g)$.

One can explicitly work out the remaining $O(g^2)$-consistency conditions~(\ref{new3.2})--(\ref{new3.3})
to find that they pose no additional constraints. It is however easy to be convinced of this fact even without
doing this exercise. To see this, let us make the identification:
\beq \k=g\l+\mathcal O(g^2),\eeq{2nd12}
and write down the first-order deformations of the master action. At $agh=0$, the result obtained from
Eq.~(\ref{2nd1}) coincides with the cubic vertices of the Aragone-Deser spin-5/2 hypergravity~\cite{Aragone:1983sz}.
The latter theory does exist and is consistent up to all orders. It follows that the cubic couplings
of our deformed theory cannot be constrained any further.

Note that the identification~(\ref{2nd12}) reproduces not only the cubic vertices but also the hypersymmetry
transformations of the theory. Indeed, one can collect the $agh=1$ terms in the master action at the zeroth
order~(\ref{freemasteraction5/2}) and at the first order~(\ref{2nd1.1}), and take functional derivatives
w.r.t.~the antifields $h^{*\m\n}$ and $\bar\ps^{*\mu\a}$ respectively, and obtain:
\bea \d_\e h_{\m\n}&=&\tfrac{3}{4}\k\,\bar\e_\a\g_{(\m}\ps_{\n)}{}^\a+\mathcal{O}\left(\k^2\right),\label{2nd13}\\
\d_\e \ps_\m{}^\a&=&\de_\m\e^\a-\tfrac{3}{8}\k\,\mathfrak{h}_{\r\s\Vert\m}
\left(\g^{\r\s}\d^\a_\b-\tfrac{2}{3}\g_\b{}^{\r\s}\g^\a\right)\e^\b+\mathcal{O}\left(\k^2\right).\eea{2nd14}
This is exactly the hypersymmetry transformations~(\ref{exp15})--(\ref{exp16}) for $n=1$ without flavor multiplicity.
In writing the above, we have omitted the gauge variations involving the ghost $C_\m$ because they just
comprise transformations under diffeomorphism.

\subsection{Uniqueness to All Orders}\label{subsec:allorder}

Now we are going to show that our results exclusively reproduce the spin-5/2 Aragone-Deser
hypergravity~\cite{Aragone:1983sz} in the metric-like version. In the context of our theory, we will
essentially repeat the arguments presented in~\cite{Boulanger:2000rq,Boulanger:2018fei}. Because
the Aragone-Deser theory has already been reproduced up to $\mathcal{O}(g)$, the deformed master
action~(\ref{brst5}) reduces to:
\beq S=S_0+gS_1^H(\hat\l)+g^2S_2+g^3S_3+\cdots,\eeq{all1}
where the superscript ``$H$'' pertains to the Aragone-Deser theory, and $\hat\l=\{\l, \b\}$
collectively denotes the coupling constants obtained at the first order in the deformation parameter $g$,
subject to the constraint~(\ref{2nd10}) imposed by second-order consistency.

Because the functional $S_2$ obeys Eq.~(\ref{brst2nd}), it must be of the form:
\beq S_2=S_2^H(\hat\l^2)+S_2',\qquad \text{with}\quad sS_2'=0.\eeq{all2}
In other words, the second-order deformation of the master action may differ with that of the Aragone-Deser
theory only by a term which belongs to the cohomology of $s$. Note that in deriving the first-order deformations
$S_1$ we did not assume anything more than locality, parity and Poincar\'e invariance. Therefore, given
Eq.~(\ref{brst1st}), it is clear that under the assumptions stated above any nontrivial element in $H(s)$ must
be the same as $S_1$, modulo coupling constants. In particular, we will have
\beq S_2'=S_1^H(\hat\l'),\qquad \text{with some}\quad \hat\l'=\{\l', \b'\}.\eeq{all3}
Next, we consider the third-order consistency condition, which reads
\beq \left(S_0, S_3\right)=-\left(S_2, S_1^H(\hat\l)\right)\quad\Rightarrow
\quad sS_3=-\left(S_2^H(\hat\l^2), S_1^H(\hat\l)\right)-\left(S_1^H(\hat\l'), S_1^H(\hat\l)\right).\eeq{all4}
Then, $S_3$ must be of the form:
\beq S_3=S_3^H(\hat\l^3)+S_3',\qquad \text{with}\quad sS_3'=-\left(S_1^H(\hat\l'), S_1^H(\hat\l)\right).\eeq{all5}
In view of Eq.~(\ref{brst2nd}), it is easy to see that the general solution for $S_3'$ will be:
\beq S_3'=2S_2^H(\hat\l\hat\l')+S_3'',\qquad \text{with}\quad sS_3''=0,\eeq{all6}
where one will again have:
\beq S_3''=S_1^H(\hat\l''),\qquad \text{with some}\quad \hat\l''=\{\l'', \b''\}.\eeq{all7}
Plugging Eqs~(\ref{all2})--(\ref{all7}) into the deformation expansion~(\ref{all1}) one then obtains:
\beq S=S_0+gS_1^H(\hat\l+g\hat\l'+g^2\hat\l'')+g^2S_2^H(\hat\l^2+2g\hat\l\hat\l')+g^3S_3^H(\hat\l^3)
+\mathcal{O}(g^4).\eeq{all8}

Now it is clear from the discussion of Section~\ref{subsec:1-def} that, to the relevant order in $g$,
the first-order deformed coupling constants $\hat\l_1\equiv \hat\l+g\hat\l'$ will be subject to the same second-order
consistency condition as the undeformed ones $\hat\l=\{\l, \b\}$, namely Eq.~(\ref{2nd10}).
Continuing to higher orders in this way, we obtain the general solution for the all-order deformed master action:
\beq S=S_0+\sum_{n=1}^{\infty}g^nS_n^H(\hat\l_\infty^n),\qquad \text{with}\quad
\hat\l_\infty\equiv\hat\l+g\hat\l'+g^2\hat\l''+\cdots.\eeq{all9}
Again, the all-order deformed coupling constants $\hat\l_\infty=\{\l_\infty, \b_\infty\}$ are subject to the
constraint~(\ref{2nd10}). In accordance with Eq.~(\ref{2nd11}), one must have $\b_\infty=-\tfrac{3}{16}\l_\infty$
in order for the fermion to couple to gravity. Given Eq.~(\ref{all9}), then the identification:
\beq \k=g\l_\infty,\eeq{all10}
proves the uniqueness of the spin-5/2 Aragone-Deser hypergravity~\cite{Aragone:1983sz} up to all orders.

\section{Uniqueness of Generalized Hypergravity}\label{sec:GenH}

Now we will generalize the analysis of Section~\ref{sec:Metric} to an arbitrary-spin fermion with
an arbitrary number of flavors. In the metric-like formulation, we will consider consistent
flat-space interactions of a spin-2 gauge field and a massless Majorana fermion of spin $s=n+3/2$,
under the assumptions of locality, parity and Poincar\'e invariance. The analysis of this section
can be copied almost verbatim from Section~\ref{sec:Metric}. We will therefore present only the
main results, skipping the detailed computations unless otherwise required.

\subsubsection*{The Free Theory}

We start with the free theory defined by Eqs.~(\ref{exp5})--(\ref{exp5.001}), which consists of
a spin-2 gauge field $h_{\m\n}$ and a massless Majorana fermion $\ps_{\m\a(n),\,I}$ with spin
$s=n+3/2$ and $N$ flavors denoted by the index $I=1,2,\ldots,N$. The set of fields and antifields are:
\beq \Phi^A=\{h_{\m\n}, C_\m, \ps_{\m\a(n),\,I}, \xi_{\a(n),\,I}\},\qquad
\Phi^*_{A}=\{h^{*\m\n}, C^{*\m}, \bar{\ps}^{*\m\a(n),\,I},\bar{\xi}^{*\a(n),\,I}\}.\eeq{gfieldsset}
The fermionic (anti)field and (anti)ghost are symmetric and $\g$-traceless in the $\a$-indices.
The fermionic gauge field $\ps_{\m\a(n),\,I}$ as such is not a Fronsdal tensor-spinor, but a
``dreibein-like fermion'' with covariant indices ($\a$-indices descend from the frame
indices through contractions with flat-space dreibein). The two descriptions are however
equivalent~\cite{Rahman:2017cxk}.

The free master action $S_0$ is given by:
\beq S_0=\int d^3x\Big(\tfrac{1}{2}h_{\m\n}\mathcal G^{\m\n}-\tf{1}{2}\bar{\ps}_{\m\a(n),I}\mathcal{R}^{\m\a(n),\,I}
+2h^{*\m\n}\de_\m C_\n -\bar{\ps}^{*\m\a(n),\,I}\de_\m\xi_{\a(n),\,I}\Big),\eeq{gfreemasteraction}
which satisfies the free master equation: $(S_0,S_0)=0$, given the properties of the various fields and antifields
summarized in Table 2 (with fermion flavor index suppressed).
\begin{table}[ht]
\caption{Properties of the Fields \& Antifields ($s=n+3/2$)}
\vspace{6pt}
\centering
\begin{tabular}{c c c c c c c c}
\hline\hline
$Z$ &$\G(Z)$~~~&~~~$\D(Z)$~~~&$pgh(Z)$ &$agh(Z)$ &$gh(Z)$ &$\epsilon(Z)$ &$\beta(Z)$\\ [0.5ex]
\hline
$h_{\m\n}$ & $2\de_{(\m} C_{\n)}$ & 0 & 0 & 0 & 0 & 0 &-\\
$C_\m$ & 0 & 0 & 1 & 0 & 1 & 1 &-\\
$h^{*\m\n}$ & 0 & $\mathcal G^{\m\n}$ & 0 & 1 & $-1$ & 1 &-\\
$C^{*\m}$ & 0 & $-2\de_\n h^{*\m\n}$ & 0 & 2 & $-2$ & 0 &-\\ \hline
$\ps_{\m\a(n)}$ & $\de_\m\xi_{\a(n)}$ & 0 & 0 & 0 & 0 & 1 &$+1$\\
$\xi_{\a(n)}$ & 0 & 0 & 1 & 0 & 1 & 0 &$+1$\\
$\bar{\ps}^{*\m\a(n)}$ & 0 & $-\bar{\mathcal R}^{\m\a(n)}$ & 0 & 1 & $-1$ & 0 &$-1$\\
$\bar{\xi}^{*\a(n)}$ & 0 & $\de_\m\bar{\ps}^{*\m\a(n)}$ & 0 & 2 & $-2$ & 1 &$-1$\\
\hline\hline
\end{tabular}
\end{table}

\subsubsection*{First-Order Deformations}

Let us note that the free master action~(\ref{gfreemasteraction}) enjoys an $O(N)$ global symmetry in the flavor space.
In addition to locality, parity and Poincar\'e invariance, we will make the assumption that the $O(N)$ symmetry is
preserved by the interactions. The complete set of first-order deformations of the master action then turns out to be:
\bea a_2&=&\,\l C_\m^*C_\n\de^\m C^\n+\b\left[2\bar\xi_{\a(n),\,I}\g^{\m\n}\xi^{*\a(n),\,I}\mathfrak{C}_{\m\n}
-\bar\xi_{\a(n),\,I}\g^\m\xi^{\a(n),\,I} C^*_\m\right],\label{g2nd1}\\
a_1&=&\l h^*_{\m\n}C_\r\left(\de^\r h^{\m\n}-\de^\m h^{\n\r}-\de^\n h^{\m\r}\right)+2\b\bar\ps_{\r\a(n),\,I}\g^{\m\n}\ps^{*\r\a(n),\,I}\mathfrak{C}_{\m\n}\nonumber\\
&&+2\b\left[2\bar\xi_{\a(n),\,I}\g^\m\ps^{\n\a(n),\,I} h^*_{\m\n}+\bar\xi_{\a(n),\,I}\g^{\m\n}
\ps^{*\r\a(n),\,I}\mathfrak{h}_{\m\n\Vert\r}\right],\label{g2nd1.1}\\
a_0&=&\l\left(\k^{-1}\mathcal{L}_{hhh}\right)+\b\,\bar{\ps}_{\m\a(n),\,I}\,\g^{\m\n\r}\g^{\s\l}\,
\ps_\n{}^{\a(n),\,I}\,\mathfrak{h}_{\s\l\Vert\,\r}\,,\eea{g2nd1.2}
where $\l$ and $\b$ are two yet-arbitrary coupling constants, and $\mathcal{L}_{hhh}$ is given by Eq.~(\ref{self03}).

Given the spin-5/2 results of Section~\ref{subsec:5half-1st}, it is easy to be convinced of
Eqs.~(\ref{g2nd1})--(\ref{g2nd1.2}) that hold for an arbitrary-spin fermion with flavor. The two theories differ
only in the fermion sector, and up to $\mathcal{O}(g)$ they map to each other by the straightforward index
replacement: $\a\leftrightarrow\{\a(n),I\}$.
This mapping becomes clear upon a comparison between the respective free master actions~(\ref{freemasteraction5/2}) and~(\ref{gfreemasteraction}), and the property tables: Table 1 and Table 2. Now that the first-order deformations
are given by the cohomology of the \emph{free} BRST differential $s$, in accordance with Eq.~(\ref{brst1st}),
clearly the mapping would continue to hold up to $\mathcal{O}(g)$ under the assumptions stated above.

\subsubsection*{Consistency at Second Order}

At the second order in the deformation parameter $g$, we compute the quantity $c_2$ given through
Eq.~(\ref{new2.1}), and cast it into the schematic expression~(\ref{2nd3}). The term $c_2^{(\l\l)}$ will
again be given by Eqs.~(\ref{2nd4.1}) and~(\ref{2nd5})--(\ref{2nd6}). The other terms in $c_2$ can
also be computed easily. One of these terms is $c_2^{(\l\b)}$, which reads:
\bea c_2^{(\l\b)}&=&-\left[\de^\m\!\left(C^*_\m C_\n\right)+C^*_\m\de^\m C_\n\right]\bar\xi_{\a(n),\,I}
\g^\n\xi^{\a(n),\,I}+4\de_\m\!\left(\bar\xi_{\a(n),\,I}\g^{\m\n}\xi^{*\a(n),\,I}\right)C_\r\de_\n C^\r
\nonumber\\&\doteq&\bar\xi^*_{\a(n),\,I}\g^{\m\n}\xi^{\a(n),\,I}\mathfrak{C}_{\m\r}\mathfrak{C}_\n{}^\r
-\tfrac{1}{2}C^*_\m\bar\xi_{\a(n),\,I}\g_\n\xi^{\a(n),\,I}\mathfrak{C}^{\m\n}+\G\text{-exact},\eea{g2nd4.2}
where the second line has been derived by rewriting the first one in terms of undifferentiated antighosts
modulo total derivatives, and then using Eqs.~(\ref{Gammaaction2}) and~(\ref{g1.9}). Last but not the least,
we need to compute $c_2^{(\b\b)}$; the result is:
\bea c_2^{(\b\b)}&=&-4\left[\bar\xi^*_{\a(n),\,I}\g^{\m\n}\mathfrak{C}_{\m\n}+\bar\xi_{\a(n),\,I}
\g^\m C^*_\m\right]_{\g\text{-}t} \left[\g^{\r\s}\xi^{\a(n),\,I}\mathfrak{C}_{\r\s}\right]_{\g\text{-}t}
\nonumber\\&=&\left(\tfrac{16}{2n+1}\right)\left[\bar\xi^*_{\a(n),\,I}\g^{\m\n}\xi^{\a(n),\,I}
\mathfrak{C}_{\m\r}\mathfrak{C}_\n{}^\r-\tfrac{1}{2}C^*_\m\bar\xi_{\a(n),\,I}\g_\n\xi^{\a(n),\,I}
\mathfrak{C}^{\m\n}\right]+\G\text{-exact},\eea{g2nd4.3}
where the subscript ``$\g\text{-}t$'' stands for the $\g$-traceless part of the term under consideration,
and the detailed technical steps leading to the second line is given in Appendix~\ref{sec:details}.

The resulting $c_2$ will be a trivial element in $H^1(\G|d)$ and thereby fulfill Eq.~(\ref{new3.1}) if and
only if the couplings $\l$ and $\b$ satisfy:
\beq \b\left\{\l+\left(\tfrac{16}{2n+1}\right)\b\right\}=0.\eeq{g2nd10}
The absence of obstructions at $\mathcal O(g^2)$ therefore necessarily requires that the number of independent coupling
constants appearing at $\mathcal O(g)$ be a single one. The quadratic constraint~(\ref{g2nd10}) gives rise to the
following branches of solutions:
\bea \b=\left\{
  \begin{array}{ll}
    -\tfrac{1}{8}\left(n+\tfrac{1}{2}\right)\l, & \text{fermion coupled to gravity}\\
    0, & \text{free fermion plus gravity.}
  \end{array}
\right.\eea{g2nd11}
As before, we would be interested in the first branch, which corresponds to a non-vanishing $\b$ and therefore to
a nontrivial coupling to gravity of the higher-spin fermion at $\mathcal O(g)$.

The remaining $O(g^2)$-consistency conditions~(\ref{new3.2})--(\ref{new3.3}) would pose no more constraints
on the first-order couplings for reasons already explained in Section~\ref{subsec:1-def}. Indeed, the
identification.~(\ref{2nd12}) would reproduce the cubic vertices~(\ref{exp6})--(\ref{exp10}) as well as
the hypersymmetry transformations~(\ref{exp15})--(\ref{exp16}) of generalized hypergravity of
Section~\ref{sec:M-H}.

\subsubsection*{Uniqueness to All Orders}

Because the generalized hypergravity theory of Section~\ref{sec:M-H} has already been reproduced up to $\mathcal{O}(g)$,
the deformed master action~(\ref{brst5}) reduces to the form:
\beq S=S_0+gS_1^{GH}(\hat\l)+g^2S_2+g^3S_3+\cdots,\eeq{gall1}
where the superscript ``$GH$'' pertains to generalized hypergravity, and $\hat\l=\{\l, \b\}$
collectively denotes the coupling constants obtained at the first order in the deformation parameter $g$,
subject to the constraint~(\ref{g2nd10}) imposed by second-order consistency.

One can repeat the logical steps of Section~\ref{subsec:allorder} to obtain the general solution for the
all-order deformed master action:
\beq S=S_0+\sum_{n=1}^{\infty}g^nS_n^{GH}(\hat\l_\infty^n),\qquad \text{with}\quad
\hat\l_\infty\equiv\hat\l+g\hat\l'+g^2\hat\l''+\cdots,\eeq{gall9}
where the all-order deformed coupling constants $\hat\l_\infty=\{\l_\infty, \b_\infty\}$ are subject to the
constraint~(\ref{g2nd10}). In accordance with Eq.~(\ref{g2nd11}), one must have
$\b_\infty=-\tfrac{1}{8}\left(n+\tfrac{1}{2}\right)\l_\infty$ in order for the higher-spin fermion to couple
to gravity. Given Eq.~(\ref{gall9}), then the identification:
\beq \k=g\l_\infty,\eeq{gall10}
proves the all-order uniqueness of the generalized hypergravity theory of Section~\ref{sec:M-H}.

\newpage
\section{Hypergravity with Cosmological Constant}\label{sec:CC}

In this section we study the consequences of a cosmological term in order to shed light on the nature of
(generalized) hypergravity in (Anti-)de Sitter space. Let us recall that, in Sections~\ref{sec:Metric}
and~\ref{sec:GenH}, while solving for the cohomology of the free BRST differential $s$ at zero ghost number,
we restricted ourselves to local functionals that are at least cubic in the fields. Relaxing this requirement
would invoke tadpole terms in the theory. However, when viewed as a theory of an interacting spin-2 gauge
field around flat space, General Relativity with(out) a cosmological constant does (not) contain tadpole
terms~\cite{Boulanger:2000rq}. Therefore, in order to admit a cosmological term we must relax the
aforementioned requirement.

\subsection{Obstruction of Spin-5/2 Hypergravity}\label{subsec:CC1}

We start with the free system of Section~\ref{subsec:5half-free}, namely that of a spin-2 and a Majorana
spin-5/2 gauge fields in flat space. In this case, the first order deformations would contain those of
Section~\ref{subsec:5half-1st} plus all possible tadpole terms. The latter kind of terms may be present
as ambiguities in the gauge symmetry deformations corresponding to non-Abelian vertices and/or in the
form of Abelian vertices. Because the consistency of the non-Abelian vertices constructed in
Section~\ref{subsec:nonAbelian} does not require ambiguities $\tilde a_1$, it suffices to reconsider
the (non-)existence of Abelian vertices discussed in Section~\ref{subsec:abelian}.

As alluded in Appendix~\ref{sec:cohomology}, under the weaker assumptions of this section,
Eq.~(\ref{noa2strict}) may admit nontrivial solutions: $a_1\neq\D\text{-exact}$.
In this case, $a_1$ will be a quadratic term with one antifield and one ghost.
The only possibility of this type is:
\beq a'_1=-\m\,\bar\xi^\a\pss^*_\a,\eeq{cc0}
where $\m$ is an arbitrary coupling constant. Note that the bosonic pair $\left\{h^*_{\m\n}, C_\m\right\}$
does not contribute here, since it is impossible to construct out of them a nontrivial Lorentz scalar.
It is easy to see that the gauge symmetry deformation~(\ref{cc0}) has a lift to an Abelian vertex through
Eq.~(\ref{cocycle3}). The result is a mass deformation of the fermion Lagrangian~\cite{Boulanger:2001wq}:
\beq a'_0=-\tfrac{1}{2}\m\,\bar\ps_{\m\a}\g^{\m\n}\ps_\n{}^\a.\eeq{cc0.1}

It remains to consider Abelian vertices that do not deform the gauge transformations and satisfy
Eq.~(\ref{noa2strict3}). The unique possibility is the linearized cosmological term~\cite{Boulanger:2000rq}:
\beq a''_0=-2\a h_{\m\n}\h^{\m\n},\eeq{cc0.2}
where $\a$ is yet another coupling constant. Along the line of discussion of Section ~\ref{subsec:abelian},
this ``vertex'' can be understood as a graviton field contracted with the current: $T^{\m\n}=-2\a\h^{\m\n}$,
which is actually a constant and therefore conserved. This current bypasses the no-go result of
Section ~\ref{subsec:abelian} in that the general solution reported in Eq.~(\ref{noa2strict5.2})
assumes that $T^{\m\n}$ is space-time dependent. No fermionic current $\Th^{\m\a}$ of this kind
exists\footnote{The only possibility $\Th^{\m\a}=\g^{\m\a}+2\h^{\m\a}$ is ruled out by the assumption of
Poincar\'e invariance.}.

Having obtained all the possible tadpole terms, we are not ready to summarize the first-order deformations
of the master action. They are given by the results~(\ref{cc0})--(\ref{cc0.2}) added to the
deformations~(\ref{2nd1})--(\ref{2nd1.2}) that exclude tadpole terms. That is,
\beq a_2=a_2^H,\qquad a_1=a_1^H-\m\,\bar\xi^\a\pss^*_\a,\qquad
a_0=a_0^H-\tfrac{1}{2}\m\,\bar\ps_{\m\a}\g^{\m\n}\ps_\n{}^\a-2\a h_{\m\n}\h^{\m\n},\eeq{cc1}
where the superscript ``$H$'' refers to the corresponding quantity in the absence of tadpole terms.
In particular, $a_2^H$, $a_1^H$ and $a_0^H$ are given by Eqs.~(\ref{2nd1})--(\ref{2nd1.2}) respectively.

Next, we would like to explore the consistency conditions~(\ref{new3.1})--(\ref{new3.3}) required
at the second order in the deformation parameter. With this end in view, we would like to compute the
quantities $c_2$, $c_1$ and $c_0$ from the first-order deformations~(\ref{cc1}) through the defining
equations~(\ref{new2.1})--(\ref{new2.3}). The details of these technically-rather-involved computations
are given in Appendix~\ref{sec:details1}. The final results are:
\beq c_2=c_2^H,\qquad c_1\approx c_1^H+\tfrac{64}{3}\,\m\b\,\de_{(\m}\bar\ps^*_{\n)}{}^{\,\m}_{s\text{-}t}
\,\Cs\,\xi^\n,\qquad c_0\approx c_0^H+\left(\tfrac{2}{3}\m^2-8\a\b\right)\bar\xi_\a\pss^\a,\eeq{cc2}
where the symbol ``$\approx$'' means equality up to $\G$-exact terms modulo $\D$-exact terms modulo
total derivatives, and the subscript ``$s\text{-}t$'' stands for a symmetric-traceless projection.
The explicit terms $\de_{(\m}\bar\ps^*_{\n)}{}^{\,\m}_{s\text{-}t}\,\Cs\xi^\n$ and $\bar\xi_\a\pss^\a$
appearing in Eqs.~(\ref{cc2}) are nontrivial elements in $H^1(\G|\D|d)$.
The second-order consistency conditions~(\ref{new3.1})--(\ref{new3.3}) then require that:
\beq \b\left(\l+\tfrac{16}{3}\b\right)=0,\qquad \m\b=0,\qquad \m^2-12\a\b=0,\eeq{cc3}
where the first relation results from requiring that $c_2=c_2^H$ be $\G$-exact modulo $d$, and so
it is identical to Eq.~(\ref{2nd10}). The terms $c_1^H$ and $c_0^H$ may also contain nontrivial elements
of $H^1(\G|\D|d)$, but they require no additional relations as already explained in
Section~\ref{subsec:1-def}.

A cosmological term in the theory sets $\a\neq0$, in which case a non-vanishing coupling of the
fermion to gravity ($\beta\neq0$) would lead us to the following relations:
\beq \b=-\tfrac{3}{16}\l\neq0,\qquad \m\b=0,\qquad \m^2=12\a\b\neq0.\eeq{cc4}
This set of mutually-incompatible relations admits no solutions for the first-order coupling
constants, and signals a cohomological obstruction.
The conclusion is that $\a$ and $\b$ cannot be simultaneously nonzero under the assumptions
of our analysis. In other words, in the presence of a cosmological constant there is no consistent
parity and Poincar\'e invariant interacting \emph{local} theory of a spin-2 and a Majorana spin-5/2
gauge fields.

\subsection{Obstruction of Generalized Hypergravity}\label{subsec:CC2}

Now we will generalize the results of the previous section to a fermion with arbitrary spin $s=n+3/2$
and an arbitrary number of flavors. We have the free system of Section~\ref{sec:GenH} and the corresponding
first-order deformations~(\ref{g2nd1})--(\ref{g2nd1.2}). When tadpole terms are included, the result
is a straightforward generalization of Eqs.~(\ref{cc1}), namely
\beq a_2=a_2^{GH},\quad a_1=a_1^{GH}-\m\bar\xi^{\a(n),\,I}\pss^*_{\a(n),\,I},\quad
a_0=a_0^{GH}-\tfrac{1}{2}\m\bar\ps_{\m\a(n),\,I}\g^{\m\n}\ps_\n{}^{\a(n),\,I}-2\a h_{\m\n}\h^{\m\n},\eeq{gcc1}
where the superscript ``$GH$'' refers to the corresponding quantity in generalized hypergravity with
tadpole terms excluded.

To check the second-order consistency conditions~(\ref{new3.1})--(\ref{new3.3}), we need to compute $c_2$,
$c_1$ and $c_0$ from the first-order deformations~(\ref{gcc1}) through Eqs.~(\ref{new2.1})--(\ref{new2.3}).
With the details given in Appendix~\ref{sec:details1}, the computations lead us to the results:
\beq c_2=c_2^{GH},\quad c_1\approx c_1^{GH}+\left(\tfrac{64n}{2n+1}\right)\m\b\,\de_{(\m}\bar\ps^*_{\n)}
{}^{\,\m,\,I}_{s\text{-}t}\,\Cs\xi^\n{}_I,\quad c_0\approx c_0^{GH}
+\left(\tfrac{2\m^2}{2n+1}-8\a\b\right)\bar\xi_{\a,\,I}\pss^{\a,\,I},\eeq{gcc2}
where again ``$\approx$'' corresponds to equality up to $\G$-exact terms modulo $\D$-exact
terms modulo total derivatives, and the subscript ``$s\text{-}t$'' means a symmetric-traceless projection.
Then, the second-order consistency conditions~(\ref{new3.1})--(\ref{new3.3}) demand that:
\beq \b\left\{\l+\left(\tfrac{16}{2n+1}\right)\b\right\}=0,\qquad n\m\b=0,\qquad \m^2
-8\left(n+\tfrac{1}{2}\right)\a\b=0,\eeq{gcc3}
where the rank $n=s-3/2$ takes the values $0,1,2,\ldots,\infty$.

Aragone-Deser hypergravity and its higher-spin generalizations correspond to $n>0$, for which there is
a cohomological obstruction if $\a\neq0$ and $\b\neq0$. In this case, the system of equations~(\ref{gcc3})
reduces to the following form:
\beq \b=-\tfrac{1}{8}\left(n+\tfrac{1}{2}\right)\l\neq0,\qquad \m\b=0,\qquad
\m^2=8\left(n+\tfrac{1}{2}\right)\a\b\neq0,\eeq{gcc4}
which admits no solutions. The conclusion is that in the presence of a cosmological constant there is no consistent
parity- and Poincar\'e-invariant interacting \emph{local} theory of a spin-2 gauge field and a massless
Majorana fermion with spin $s\geq5/2$.

As a byproduct of our analysis it follows that the second-order obstruction disappears for extended supergravity
theories in $D=3$. To see this, note that supergravity corresponds to $n=0$, which admits the
the following solution of the system of equations~(\ref{gcc3}):
\beq \b=-\tfrac{1}{16}\l\neq0,\qquad \m^2=4\a\b\neq0.\eeq{gcc5}
These requirements are indeed mutually compatible. This is hardly a surprise given that three-dimensional extended
Anti-de Sitter supergravity does exist as a consistent theory of a spin-2 and a number of spin-3/2 gauge
fields~\cite{Achucarro:1987vz}. It is however reassuring that our analysis is in agreement with
this well-known fact.

\section{Remarks}\label{sec:remarks}

In this article we have studied consistent interactions of a spin-2 gauge field and a higher-spin massless Majorana fermion
in 3D flat space. Under the assumptions of locality, parity and Poincar\'e invariance, we have found in the
metric-like formulation that the interacting theory is unique: Aragone-Deser hypergravity~\cite{Aragone:1983sz} or
a higher-spin generalization thereof. Local hypersymmetry and its higher-spin counterparts follow automatically as the only
consistent deformations of the gauge transformations. We also showed that an extension of (generalized) hypergravity to an
arbitrary number of fermion flavors does not change these features. In the presence of a cosmological constant, however, there
is no consistent theory of a higher-spin fermionic gauge field coupled to gravity under those assumptions.

One possible way to generalize our analysis is to relax the assumption of parity. After all, parity-odd cubic vertices do
exist in $D=3$, at least for bosons~\cite{Kessel:2018ugi}. This would give rise to an additional graviton self coupling at
the cubic level. On the other hand, our results for cubic fermion-graviton-fermion cross couplings$-$non-Abelian or
Abelian$-$did not rely on parity. Therefore, our analysis excludes the possibility of parity-odd cubic vertices (of the
type $2-s-s$) between a graviton and a higher-spin Majorana fermion. However, it could still be possible to construct such
vertices by endowing the graviton with a Chan-Paton factor as in colored gravity~\cite{Gwak:2015vfb,Gwak:2015jdo}.

We would like to emphasize that our no-go results for cosmological hypergravity and its higher-spin generalizations
no longer hold if locality is given up. Non-locality, if necessary, might be intrinsic and/or result from having
integrated out additional degrees of freedom in the consistent interacting theory. Indeed, it is the second
possibility that gives way to the yes-go results of~\cite{Zinoviev:2014sza}, where it was shown that in Anti-de
Sitter space the inclusion of a massless spin-4 field makes hypergravity consistent (see also~\cite{Henneaux:2015tar}).
For a fermion of arbitrary spin $s=n+3/2$, the yes-go works with the even spins $2, 4,\ldots, 2n+2$. In the context of
our analysis, it would be interesting to see how the inclusion of these additional fields in the spectrum remove the
otherwise-present cohomological obstructions.

The coupling of hypergravity to matter could be interesting. In flat space, supergravity and generalized hypergravity
are analogous in that they both contain only two gauge fields: a graviton and a Majorana fermion.
This is in sharp contrast with 3D bosonic higher-spin gauge theories, where starting from
spin 2 one would need \emph{all} the (even) integer spins up to some higher value. It is however expected
that matter coupling of a bosonic theory would necessarily call for an infinite tower of higher-spin
fields. For generalized hypergravity, it is plausible that coupling to a suitable matter multiplet is consistent
without invoking extra higher-spin gauge fields. This expectation arises from the matter coupling of $D=3$
supergravity~\cite{Kuzenko:2011xg}. One could also study models with spontaneously broken hypersymmetry 
(see, for example,~\cite{Bansal:2018qyz}). We leave these as future work.

\subsection*{Acknowledgments}

The author is grateful to M.~Henneuax and G.~Lucena G\'omez, who were closely involved in this work at the initial stage.
He would like to thank N.~Boulanger, A.~Campoleoni, D.~Chow and A.~Leonard for useful discussions and valuable comments.
RR acknowledges the kind hospitality and support of the Erwin Schr\"odinger International Institute
for Mathematics and Physics and the organizers of the scientific activity ``Higher Spins and Holography''
(March 11--April 05, 2019), during which this work has been completed.

\begin{appendix}
\numberwithin{equation}{section}

\section{Curvatures, Identities \& EoMs}\label{sec:curvatures}

In this appendix we discuss some important properties of the curvatures and curls of the different
fields under consideration. We also write down the various forms of the equations of motion (EoM)
in terms of these objects, which would help us identify $\D$-exact terms.

\subsubsection*{Spin-2 Gauge Field}

The spin-2 graviton field is denoted by $h_{\m\n}$, while its 1-curl by a Fraktur letter: $\mathfrak{h}_{\m\n\Vert\r}$. The 1-curl is antisymmetric
in its first two indices, and obeys the Bianchi identities:
\beq \mathfrak{h}_{[\m\n\Vert\r]}=0\quad\Leftrightarrow\quad\mathfrak{h}_{\m[\n\Vert\r]}=-\tf{1}{2}\mathfrak{h}_{\n\r\Vert\m},\qquad
\de_{[\m}\mathfrak{h}_{\n\r]\Vert\s}=0.\eeq{x1}
The 2-curl of the graviton is simply the linearized Riemann tensor: $R_{\m\n}{}^{\r\s}\equiv 4\de_{[\m}\de^{[\r}h_{\n]}{}^{\s]}$.
It has the symmetry properties: $R_{\m\n\r\s}=R_{[\m\n][\r\s]}=R_{\r\s\m\n}$, and obeys the Bianchi identities:
\beq R_{[\m\n\r]\s}=0, \qquad \de_{[\m}R_{\n\r]\a\b}=0.\eeq{x2}
The linearized Ricci tensor is defined as $R_{\m\n}\equiv R_{\m\r\n}{}^\r$, whose trace in turn gives the linearized Ricci scalar $R\equiv R^\m_\m$.
It is clear that the quantity $\mathcal G_{\m\n}$ appearing in Eq.~(\ref{3.1}) is nothing but the linearized Einstein tensor.
The Euler-Lagrange EoMs are reflected in:
\beq \mathcal G_{\m\n}= R_{\m\n}-\tf{1}{2}\h_{\m\n}R=\D h^*_{\m\n}.\eeq{grav1}
In $D=3$, however, the Weyl tensor vanishes identically. One can write:
\beq R_{\m\n\r\s}=\ve_{\m\n\a}\ve_{\r\s\b}\mathcal G^{\a\b},\qquad \mathcal G^{\m\n}=\tfrac{1}{4}\ve^{\m\a\b}\ve^{\n\r\s}R_{\a\b\r\s}.\eeq{Riemann-G}
It follows immediately that, in $D=3$, the Riemann tensor itself is $\D$-exact:
\beq R_{\m\n\r\s}=\ve_{\m\n\a}\ve_{\r\s\b}\D h^{*\a\b}.\eeq{Riemann-eom}
Taking traces of the above equation, one can further write:
\beq R_{\m\n}=\D\left(h^*_{\m\n}-\h_{\m\n}h^{*\prime}\right),\qquad R=-2\D h^{*\prime},\eeq{x4}
where a ``prime'' denotes a trace, $h^{*\prime} \equiv h^{*\m}_{~\m}$. One can take a divergence of the second relation in Eq.~\rf{Riemann-G}, and use
the first Bianchi in Eq.~(\ref{x2}) to write:
\beq \de_\m \mathcal G^{\m\n}=\tfrac{1}{4}\ve^{\m\a\b}\ve^{\n\r\s}\de_\m R_{\a\b\r\s}=0.\eeq{grav2}

For the free spin-2 system, note that $\D^2$ vanishes trivially on the fields and antifields expect when it acts on the antighost $C^{*\m}$.
Now, one can write from Table 1 that
\beq \D^2 C^{*\m}=\D\left(-2\de_\n h^{*\m\n}\right)=-2\de_\n\mathcal G^{\m\n},\eeq{D-2nilp}
which vanishes identically because of Eq.~(\ref{grav2}), in accordance with the nilpotency of $\D$.

\subsubsection*{Higher-Spin Fermion}\label{subsec:arbitrary}

A massless Majorana fermion of spin $s=n+3/2$ is denoted by the tensor-spinor $\ps_{\m\a_1\ldots\a_n}$,
or $\ps_{\m\a(n)}$ as a short-hand notation. The field is symmetric and $\g$-traceless w.r.t.~the $\a$-indices: $\ps_{\m\a_1\ldots\a_n}=\ps_{\m(\a_1\ldots\a_n)}$, and $\g^{\a_n}\ps_{\m\a_1\ldots\a_n}=0$. For example, a
spin-5/2 field that corresponds to $n=1$ is denoted by the tensor-spinor $\ps_{\m\a}$, which is $\g$-traceless
w.r.t.~the second index: $\g^\a\ps_{\m\a}=0$.

The EoMs of the higher-spin fermion (possibly with an implicit flavor index) are:
\beq \mathcal R^{\m\a(n)}\equiv\g^{\m\n\r}\de_\n\ps_\r{}^{\a(n)}=\D\ps^{*\m\a(n)}.\eeq{eoms}
For the Majorana-conjugate spinor, one obtains
\beq \D\bar{\ps}^{*\m\a(n)}=-\bar{\mathcal R}^{\m\a(n)}\equiv\bar{\ps}_\n{}^{\a(n)}\overleftarrow{\de}_\r\g^{\m\n\r}.\eeq{eomsM}

It is easy to see that the 1-curl of the fermion field itself is $\D$-exact. One can simply contract Eqs.~(\ref{eoms})
and~(\ref{eomsM}) with $\ve_{\m\a\b}$ to obtain
\beq \de_{[\m}\ps_{\n]}{}^{\a(n)}=\D\left(\tfrac{1}{2}\ve_{\m\n\r}\ps^{*\r\a(n)}\right),\qquad
\bar{\ps}_{[\m}{}^{\a(n)}\overleftarrow{\de}_{\n]}=\D\left(\tfrac{1}{2}\ve_{\m\n\r}\bar{\ps}^{*\r\a(n)}\right),\eeq{eomsmagic}
by virtue of the relation: $\g^{\m\n\r}=-\ve^{\m\n\r}\,\mathbb{I}$. Taking $\g$-traces of Eqs.~(\ref{eomsmagic}) would lead to
\beq \ds\ps_\m{}^{\a(n)}-\de_\m\pss^{\a(n)}=\D\left(-\g_{\m\n}\bar{\ps}^{*\n\a(n)}\right),\qquad \bar{\ps}_\m{}^{\a(n)}\overleftarrow{\ds}
-\bar{\pss}^{\a(n)}\overleftarrow{\de}_\m=\D\left(\bar{\ps}^{*\n\a(n)}\g_{\m\n}\right).\eeq{eomsgtr1}
Furthermore, double $\g$-traces of Eqs.~(\ref{eomsmagic}) give:
\beq \ds\pss^{\a(n)}-\de_\m\ps^{\m\a(n)}=\D\pss^{*\a(n)},\qquad\bar{\pss}^{\a(n)}\overleftarrow{\ds}
-\bar{\ps}^{\m\a(n)}\overleftarrow{\de}_\m=\D\bar{\pss}^{*\a(n)}.\eeq{eomsgtr2}
Another form of the EoMs can be obtained by taking a trace of Eq.~(\ref{eomsmagic}):
\begin{equation}\label{eomsgtr2.5}
\begin{split}
\de_\n\ps_\m{}^{\n\a(n-1)}-\de_\m\ps^{\prime\a(n-1)}=\ve_{\m\n\r}\D\ps^{*\n\r\a(n-1)},\\
\bar\ps_\m{}^{\n\a(n-1)}\overleftarrow{\de}_\n-\bar\ps^{\prime\a(n-1)}\overleftarrow{\de}_\m
=\ve_{\m\n\r}\D\bar\ps^{*\n\r\a(n-1)}.
\end{split}
\end{equation}
Finally, a $\g$-trace of the above equations gives:
\begin{equation}\label{eomsgtr2.6}
\begin{split}
\de_\m\pss^{\m\a(n-1)}-\ds\ps^{\prime\a(n-1)}=\g_{\m\n}\D\ps^{*\m\n\a(n-1)},\\
\bar\pss^{\m\a(n-1)}\overleftarrow{\de}_\m-\bar\ps^{\prime\a(n-1)}\overleftarrow{\ds}
=\D\bar\ps^{*\m\n\a(n-1)}\g_{\m\n}.\end{split}\end{equation}

For the free fermion, $\D^2$ vanishes trivially on the fields and antifields expect when it acts on the antighost $\xi^{*\a(n)}$.
Let us note from Table 2 that
\beq \D\bar{\xi}^{*\a(n)}=\bar{\psi}^{*\m\a(n)}\overleftarrow{\de}_\m,\qquad \D\xi^{*\a(n)}=\de_\m\ps^{*\m\a(n)}.\eeq{D-xis}
Therefore, one can write from Eqs.~(\ref{eoms})--(\ref{eomsM}),
\beq \D^2\bar{\xi}^{*\a(n)}=-\bar{\mathcal R}^{*\m\a(n)}\overleftarrow{\de}_\m,\qquad \D^2\xi^{*\a(n)}=\de_\m\mathcal R^{\m\a(n)},\eeq{D-2nilpxis}
which vanish identically, in accordance with the nilpotency of $\D$.

\section{The Cohomology of $\G$}\label{sec:cohomology}

In this appendix we clarify and prove some important facts about the cohomology of $\G$, used throughout the main
text. Let us recall that the action of $\G$ is given by:
\bea &\G h_{\m\n}=2\de_{(\m} C_{\n)},&\label{Gammaaction1}\\&\G\ps_{\m\a(n)}=\de_\m\xi_{\a(n)},
\qquad \G\bar\ps_{\m\a(n)}=\de_\m\bar\xi_{\a(n)}.&\eea{Gammaaction2}
The nontrivial elements in $H(\G)$ are those gauge-invariant objects that cannot be written as gauge variations of
some other objects. Let us enumerate and discuss briefly about the various types of such elements, and write down
some important $\G$-exact objects.
\vspace{5pt}
\newline
\noindent{\bf Curvatures:} The curvatures $\{R_{\m\n\r\s}, \de_{[\m}\ps_{\n]\a(n)}\}$ and derivatives thereof are
in $H(\G)$. The $\G$-closure of the curvatures is easy to see. For the linearized Riemann tensor, $R_{\m\n\r\s}$,
one can take a 2-curl of Eq.~\rf{Gammaaction1} and use the commutativity of partial derivatives to find:
\beq \G R_{\m\n}{}^{\r\s}=\G\left(4\de^{[\r}\de_{[\m}h_{\n]}{}^{\s]}\right)=4\de^{[\r}\de_{[\m}\de_{\n]}C^{\s]}
+4\de^{[\r}\de_{[\m}\de^{\s]}C_{\n]}=0.\eeq{g1}
One can also take a 1-curl of the first equation of~\rf{Gammaaction2} to obtain:
\beq \G\,\de_{[\m}\ps_{\n]}{}^{\a(n)}=0,\eeq{g-ncurl}
and similarly for the Majorana conjugate. The fact that the curvature $\de_{[\m}\ps_{\n]\a(n)}$ is $\G$-closed
holds good irrespective any $\g$-trace constraint on the fermionic ghost.
\newpage
The curvatures have $pgh=0$, and therefore cannot be $\G$-exact since the latter kind of objects must have $pgh>0$.
It follows that the curvatures are nontrivial elements in the cohomology of $\G$. So are the derivatives of the curvatures.
\vspace{5pt}
\newline
\noindent{\bf Antifields:} The antifields  $\{h^{*\m\n}, C^{*\m}, \bar{\ps}^{*\m\a(n)}, \bar{\xi}^{*\a(n)}\}$ and
their derivatives are $\G$-closed objects since $\G$ does not act on the antifields. On the other hand, they cannot
be $\G$-exact because they have $pgh=0$. These objects therefore belong to the cohomology of $\G$.
\vspace{5pt}
\newline
\noindent{\bf Ghosts \& Ghost-Curls:} Because $\G$ does not affect them, the \emph{undifferentiated} ghosts
$\{C_\m, \xi_{\a(n)}\}$ are $\G$-closed objects. They, on the other hand, are not $\G$-exact terms since the latter
must contain at least one derivative of a ghost in accordance with Eqs.~\rf{Gammaaction1}--\rf{Gammaaction2}.

Any derivative of the bosonic ghost $C_\m$ is $\G$-closed. Some derivatives can be $\G$-exact, e.g.,
symmetrized derivatives of $C_\m$ are trivial in $H(\G)$ since by definition:
$\de_{(\m}C_{\n)}=\tf{1}{2}\G h_{\m\n}$. A 1-curl of the bosonic ghost, $\mathfrak{C}_{\m\n}\equiv2\de_{[\m}C_{\n]}$,
however is not $\G$-exact. We have:
\beq \de_\m C_\n=\de_{(\m}C_{\n)}+\de_{[\m}C_{\n]}=\tf{1}{2}\G h_{\m\n}+\tf{1}{2}\mathfrak{C}_{\m\n}\,.\eeq{g1.9}
Any derivative of $\mathfrak{C}_{\m\n}$ is however $\G$-exact. This can be seen by taking a curl of Eq.~\rf{Gammaaction1}
and using the commutativity of partial derivatives, which lead us to the result:
\beq \de_\r\mathfrak{C}_{\m\n}=\G\mathfrak{h}_{\m\n\Vert\r}\,.\eeq{ghcurl}

On the other hand, derivatives of the fermionic ghost $\xi_{\a(n)}$ are always trivial in the cohomology of $\G$. This is because the gradient of $\xi_{\a(n)}$
is $\G$-exact, as we see from Eq.~(\ref{Gammaaction2}). So, we may exclude from the cohomology of $\G$ any derivative of the fermionic ghost.

\subsubsection*{The Cohomology of $\G$ at $agh=1$ and $gh=0$}

Let us consider an element that belongs to the cohomology of $\G$ at $agh=1$ and $gh=0$. It contains \emph{only one}
of the antifields $\{h^{*\m\n}, \bar\ps^{*\m\a(n)}\}$ and \emph{only one} of the ghosts $\{C_\m, \xi_{\a(n)}\}$ in
accordance with the gradings. The fields $\{h_{\m\n}, \ps_{\m\a(n)}\}$ may or may not show up.

In order that the element under consideration belongs to $H(\G)$, any nonzero number of fields may show up only in the
form of the curvatures $\{R_{\m\n\r\s}, \de_{[\m}\ps_{\n]\a(n)}\}$. However, in $D=3$ the curvatures are $\D$-exact
(see Appendix~\ref{sec:curvatures}). Therefore, any element in $H(\G)$ containing \emph{at least} one of the original
fields at $agh=1$ and $gh=0$ must be $\D\text{-exact}$.

If none of the original fields $\{h_{\m\n}, \ps_{\m\a(n)}\}$ are present, the element under consideration will
not necessarily be $\D$-exact. This situation however falls outside the scope of our analyses in
Sections~\ref{sec:Metric} and~\ref{sec:GenH}. In this case the element is quadratic with one antifield and one ghost,
and as such its possible presence in the master action would invoke tadpole terms. Such terms are taken into account
in the analysis of Section~\ref{sec:CC}.

\section{Technical Details}\label{sec:technical}

This appendix gives some technical details we omitted in the main text for the sake of readability.
Appendix~\ref{sec:gamma} enumerates some useful $\g$-matrix identities, while Appendices~\ref{sec:details}
and~\ref{sec:details1} elaborate on the technical steps leading to some of the important derivations.

\subsection{Gamma Matrix Identities}\label{sec:gamma}

We would like to deal with flat-space $\g$-matrices: $\g_\m\equiv\g_a\bar{e}^{\,a}_\m$, where $\bar{e}^{\,a}_\m$ is the
flat-space vielbein. Below we enumerate some useful $\g$-matrix identities. For the sake of generality the identities
are written an in arbitrary number of space-time dimensions $D$. In $D=3$, some of the
identities will simplify in that the antisymmetric $4$-gamma $\g^{\m\n\r\s}$ or any other antisymmetrization
of more than three indices vanish.

We start with the symbol $\h^{\m\n|\a\b}\equiv\tfrac{1}{2}\left(\h^{\m\a}\h^{\n\b}-\h^{\m\b}\h^{\n\a}\right)$, which
is related to the antisymmetric 4-gamma in the following way:
\beq \h^{\m\n|\a\b}=\tfrac{1}{2}\g^{\m\n}\h^{\a\b}
-\tfrac{1}{4}\left(\g^\a\g^\b\g^{\m\n}+\g^{\m\n}\g^\a\g^\b\right)+\tfrac{1}{2}\g^{\m\n\a\b}.\eeq{etadefined}
An antisymmetric 3-gamma can be expanded in two ways:
\bea \g^{\m\a\b}&=&\g^\m\g^{\a\b}-2\h^{\m[\a}\g^{\b]}=\g^\m\left(\g^\a\g^\b-\h^{\a\b}\right)-2\h^{\m[\a}\g^{\b]},\label{3gamma1}\\
\g^{\a\m\n}&=&\g^{\a\m}\g^\n+2\h^{\n[\a}\g^{\m]}=\left(\g^\a\g^\m-\h^{\a\m}\right)\g^\n+2\h^{\n[\a}\g^{\m]},\eea{3gamma1.5}
whereas its $\g$-trace is given by:
\beq \g_\m\g^{\m\a\b}=(D-2)\g^{\a\b}=(D-2)\left(\g^\a\g^\b-\h^{\a\b}\right)=\g^{\a\b\m}\g_\m.\eeq{2gamma1}
The commutator of a 1-gamma and an antisymmetric 2-gamma reads:
\beq \left[\g^\a,\g^{\m\n}\right]=4\h^{\a[\m}\g^{\n]}.\eeq{commutat1}
The product of a pair of antisymmetric 2-gamma can be written as:
\beq \g^{\m\n}\g^{\r\s}\,=\,\g^{\m\n\r\s}+2\h^{\m\n|\r\s}+4\g^{[\m}\h^{\n][\r}\g^{\s]}.\eeq{commutat2}
Antisymmetrization in two indices in the above identity leads to:
\bea \g^{\r\s}\g^{\m\n}-\g^{\r\m}\g^{\s\n}&=&\h^{\r\n}\g^{\s\m}
+\g^\r\left(\g^{[\s}\g^{\m]\n}-\h^{\n[\s}\g^{\m]}\right),\label{commutat3}\\
\g^{\m\n}\g^{\r\s}-\g^{\m\s}\g^{\r\n}&=&-\h^{\r\m}\g^{\n\s}
-\left(\g^{\m[\n}\g^{\s]}+\h^{\m[\n}\g^{\s]}\right)\g^\r.\eea{commutat4}
The commutator of a 1-gamma and an antisymmetric 3-gamma reads:
\beq \left[\g^\a,\g^{\m\n\r}\right]=2\g^{\a\m\n\r}.\eeq{commutat5}
%

\subsection{Computation of $c_2^{(\b\b)}$}\label{sec:details}

First, we derive the expression~(\ref{2nd9}) for $c_2^{(\b\b)}$ starting from Eq.~(\ref{2nd4.3}).
With this end in view, let us define the following quantities:
\beq \bar{A}_\a\equiv\bar\xi^*_\a\g^{\m\n}\mathfrak{C}_{\m\n}+\bar\xi_\a\g^\m C^*_\m,\qquad B^\a\equiv\g^{\m\n}\xi^\a\mathfrak{C}_{\m\n}.\eeq{ABdefined}
Their $\g$-traces will be given by:
\beq \bar{\As}=-4\left(\bar\xi^{*\a}\g^\m\mathfrak{C}_{\a\m}+\tfrac{1}{2}\bar\xi_\a\g^{\a\m}C^*_\m\right),\qquad \Bs=4\g^\m\xi^\a\mathfrak{C}_{\a\m},\eeq{ABtrdefined}
where we have made use of the commutator~(\ref{commutat1}), given the $\g$-tracelessness of the fermionic (anti)ghost.
Then, we can rewrite Eq.~(\ref{2nd4.3}) as:
\beq c_2^{(\b\b)}=-4\left(\bar{A}_\a-\tfrac{1}{3}\,\bar\As\,\g_\a\right)\left(B^\a-\tfrac{1}{3}\g^\a\Bs\right)
=-4\left(\bar{A}_\a B^\a-\tfrac{1}{3}\,\bar\As\,\Bs\right).\eeq{2nd8}
The explicit expressions of the terms $\bar{A}_\a B^\a$ and $\bar\As\,\Bs$ are given by:
\bea &\bar{A}_\a B^\a=\left(\bar\xi^*_\a\g^{\m\n}\g^{\r\s}\xi^\a\right)\mathfrak{C}_{\m\n}\mathfrak{C}_{\r\s}
+C^*_\m\left(\bar\xi_\a\g^\m\g^{\n\r}\xi^\a\right)\mathfrak{C}_{\n\r},&\label{2nd20}\\
&\bar\As\,\Bs=-16\left(\bar\xi^{*\a}\g^\m\g^\n\xi^\b\right)\mathfrak{C}_{\a\m}\mathfrak{C}_{\b\n}
-8\,C^*_\m\left(\bar\xi_\a\g^{\a\m}\g^\n\xi^\b\right)\mathfrak{C}_{\b\n}.&\eea{2nd21}
While computing the right hand side of Eq.~(\ref{2nd20}), in the first term we use
identity~(\ref{commutat2}) and the Grassmann oddness of the bosonic ghost-curl: $\mathfrak{C}_{\m\n}\mathfrak{C}_{\r\s}=-\mathfrak{C}_{\r\s}\mathfrak{C}_{\m\n}$. In the second
term we use identity~(\ref{3gamma1}) and the vanishing of the quantity $\bar\xi_\a\g^{\m\n\r}\xi^\a$.
The result is:
\beq \bar{A}_\a B^\a=-4\left(\bar\xi^*_\a\g^{\m\n}\xi^\a\mathfrak{C}_{\m\r}\mathfrak{C}_\n{}^{\r}
-\tfrac{1}{2}C^*_\m\bar\xi_\a\g_\n\xi^\a\mathfrak{C}^{\m\n}\right).\eeq{comp1}

The right hand side of Eq.~(\ref{2nd21}) is trickier to compute. By virtue of the relation:
\beq \left(\bar\xi^{*\a}\g^{\m\n}\xi^\b\right)\mathfrak{C}_{\a\m}\mathfrak{C}_{\b\n}=0,\eeq{specialidentity}
we can cast the first term into the form:
$\left(\bar\xi^*_\a\,\h^{\a\b|\a'\b'}\xi_\b\right)\mathfrak{C}_{\a'\r}\mathfrak{C}_{\b'}{}^{\r}$,
which can then be rewritten in terms of an antisymmetric 2-gamma because of identity~(\ref{etadefined}).
The second term on the other hand reduces, on account of identity~(\ref{3gamma1.5}), to the form:
\beq C^*_\m\left(\bar\xi_\a\g^{\a\m}\g^\n\xi^\b\right)\mathfrak{C}_{\b\n}
=C^*_\m\,\bar\xi_\a\left(\g^{\a\m\n}\mathfrak{C}^\b{}_\n\right)\xi_\b
+C^*_\m\left(\bar\xi^\a\g^\m\xi^\b\right)\mathfrak{C}_{\a\b}.\eeq{comp2}
The second term on the right hand side vanishes since $\bar\xi^\a\g^\m\xi^\b$ is symmetric in $(\a,\b)$.
In the first term we can make the splitting: $\mathfrak{C}^\b{}_\n=2\de^\b C_\n+\G\text{-exact}$,
and consider the term $\g^{\a\m\n}\de^\b C_\n$, in which antisymmetry in $(\a,\b)$ is imposed by the Majorana
spinor. Since the complete antisymmetrization: $\g^{[\a\m\n}\de^{\b]} C_\n$ vanishes in $D=3$, Eq.~(\ref{comp2})
reduces to:
\bea C^*_\m\left(\bar\xi_\a\g^{\a\m}\g^\n\xi^\b\right)\mathfrak{C}_{\b\n}&=&\G\text{-exact}
-C^*_\m\,\bar\xi_\a\left(\g^{\m\a\b}\de^\n C_\n-\g^{\n\a\b}\de^\m C_\n\right)\xi_\b\nonumber\\
&=&\G\text{-exact}-\tfrac{1}{2}C^*_\m\bar\xi_\a\g_\n\xi^\a\mathfrak{C}^{\m\n},\eea{comp3}
where the second line was obtained by using Eq.~(\ref{g1.9}) and its trace, and identity~(\ref{3gamma1}).
Therefore, we have reduced Eq.~(\ref{2nd21}) to the form:
\beq \bar{\As}\,\Bs=-8\left(\bar\xi^*_\a\g^{\m\n}\xi^\a\mathfrak{C}_{\m\r}\mathfrak{C}_\n{}^{\r}
-\tfrac{1}{2}C^*_\m\bar\xi_\a\g_\n\xi^\a\mathfrak{C}^{\m\n}\right)+\G\text{-exact}.\eeq{comp4}
Now plugging Eqs.~(\ref{comp1}) and~(\ref{comp4}) into Eq.~(\ref{2nd8}) gives:
\beq c_2^{(\b\b)}=\tfrac{16}{3}\left(\bar\xi^*_\a\g^{\m\n}\xi^\a\mathfrak{C}_{\m\r}\mathfrak{C}_\n{}^{\r}
-\tfrac{1}{2}C^*_\m\bar\xi_\a\g_\n\xi^\a\mathfrak{C}^{\m\n}\right)+\G\text{-exact},\eeq{comp5}
which is precisely Eq.~(\ref{2nd9}) that we wanted to derive.

It remains to be shown that relation~(\ref{specialidentity}) indeed holds. To see this let us write:
\beq \left(\bar\xi^{*\a}\g^{\m\n}\xi^\b\right)\mathfrak{C}_{\a\m}\mathfrak{C}_{\b\n}=
-\tfrac{1}{2}\left(\bar\xi^*_\r\g^{\r\s}\g^{\m\n}\xi^\b\right)\mathfrak{C}_{\s\m}\mathfrak{C}_{\b\n}
+\tfrac{1}{2}\left(\bar\xi^{*\a}\g^{\m\n}\g^{\r\s}\xi_\r\right)\mathfrak{C}_{\a\m}\mathfrak{C}_{\s\n},
\eeq{specialid1}
where we have used the identities: $\g^{\r\s}=\g^\r\g^\s-\h^{\r\s}=\g^\s\g^\r+\h^{\r\s}$, and the
$\g$-tracelessness of the fermionic (anti)ghost. The bosonic ghost curls on the right hand side of
Eq.~(\ref{specialid1}) impose certain antisymmetry in the indices of the $\g$-matrix products, so
that identities~(\ref{commutat3})--(\ref{commutat4}) become directly applicable. Then, it follows
from the properties of the (anti)ghosts that the two terms on the right hand side of
Eq.~(\ref{specialid1}) exactly cancel each other, which proves relation~(\ref{specialidentity}).
This completes our derivation of Eq.~(\ref{comp5}). $\blacksquare$

Next, we find an expression for the quantity $c_2^{(\b\b)}$ for an arbitrary-spin fermion with flavor, i.e.,
derive the second line of Eq.~(\ref{g2nd4.3}) from the first. The derivation is quite similar to that of the
spin-5/2 case. The only subtlety is that $\g$-traceless projections of the quantities under consideration give
rise to factors that depend on the rank $n=s-3/2$. Explicitly, the first line of Eq.~(\ref{g2nd4.3}) yields:
\bea c_2^{(\b\b)}&=&-4\left[\bar{A}_{\a(n)}-\left(\tfrac{n}{2n+1}\right)\bar\As_{\a(n-1)}\,\g_\a\right]
\left[B^{\a(n)}-\left(\tfrac{n}{2n+1}\right)\g^\a\Bs^{\a(n-1)}\right]\nonumber\\
&=&-4\left[\bar{A}_{\a(n)} B^{\a(n)}-\left(\tfrac{n}{2n+1}\right)\bar\As_{\a(n-1)}\,\Bs^{\a(n-1)}\right],
\eea{g2nd8}
where we have defined:
\beq \bar{A}_{\a(n)}\equiv\bar\xi^*_{\a(n)}\g^{\m\n}\mathfrak{C}_{\m\n}+\bar\xi_{\a(n)}\g^\m C^*_\m,\qquad B^{\a(n)}\equiv\g^{\m\n}\xi^{\a(n)}\mathfrak{C}_{\m\n}\,.\eeq{gABdefined}
with the flavor index $I$ made implicit in order to avoid cumbersome notations. The computation of
$\bar{A}_{\a(n)} B^{\a(n)}$ and $\bar\As_{\a(n-1)}\,\Bs^{\a(n-1)}$ goes exactly the same way as before,
giving:
\bea &\bar{A}_{\a(n)} B^{\a(n)}=-4\left[\bar\xi^*_{\a(n)}\g^{\m\n}\xi^{\a(n)}\mathfrak{C}_{\m\r}\mathfrak{C}_\n{}^{\r}
-\tfrac{1}{2}C^*_\m\bar\xi_{\a(n)}\g_\n\xi^{\a(n)}\mathfrak{C}^{\m\n}\right],&\label{gcomp1}\\
&\bar\As_{\a(n-1)}\,\Bs^{\a(n-1)}=-8\left[\bar\xi^*_{\a(n)}\g^{\m\n}\xi^{\a(n)}\mathfrak{C}_{\m\r}\mathfrak{C}_\n{}^{\r}
-\tfrac{1}{2}C^*_\m\bar\xi_{\a(n)}\g_\n\xi^{\a(n)}\mathfrak{C}^{\m\n}\right]+\G\text{-exact}.&\eea{gcomp4}
These are straightforward generalizations of the spin-5/2 equations~(\ref{comp1})
and~(\ref{comp4}) respectively. Now plugging Eqs.~(\ref{gcomp1})--(\ref{gcomp4}) into Eq.~(\ref{g2nd8}) gives:
\beq c_2^{(\b\b)}=\left(\tfrac{16}{2n+1}\right)\left[\bar\xi^*_{\a(n)}\g^{\m\n}\xi^{\a(n)}\mathfrak{C}_{\m\r}
\mathfrak{C}_\n{}^{\r}
-\tfrac{1}{2}C^*_\m\bar\xi_{\a(n)}\g_\n\xi^{\a(n)}\mathfrak{C}^{\m\n}\right]+\G\text{-exact},\eeq{gcomp5}
which is precisely the second line of Eq.~(\ref{g2nd4.3}) with an implicit flavor index.

\subsection{Computation of $c_i$'s with Tadpole Terms}\label{sec:details1}

We start with the computation of $c_2$, $c_1$ and $c_0$ in Section~\ref{subsec:CC1}, i.e., the derivation
of Eqs.~(\ref{cc2}) from the first-order deformations~(\ref{cc1}) through
Eqs.~(\ref{new2.1})--(\ref{new2.3}). While $c_2$ is computed in Section~\ref{subsec:1-def}, in order to compute
$c_1$ let us simplify Eq.~(\ref{new2.2}) to:
\beq \int c_1=-\left(\frac{\d^R}{\d\Phi^A}\int a_1\right)\left(\frac{\d^L}{\d\Phi^*_A}\int a_2\right)
-\left(\frac{\d^R}{\d\Phi^A}\int a_1\right)\left(\frac{\d^L}{\d\Phi^*_A}\int a_1\right).\eeq{acc1}
Note that a functional derivative w.r.t.~the fermionic anti(filed) or anti(ghost)
ought to be $\g$-traceless in the $\a$-index. It is easy to see that $c_1$ takes the form:
\beq c_1=c_1^H+\m\b c_1^{(\m\b)},\eeq{acc2}
with $c_1^H$ being the hypergravity contribution without tadpole terms, and $c_1^{(\m\b)}$ given by:
\beq c_1^{(\m\b)}=2\left(\bar\ps^*_{\m\a}\g^\m\right)_{\g\text{-}t}\left(\g^{\m\n}\xi^\a\mathfrak{C}_{\m\n}\right)_{\g\text{-}t}
-2\left(\bar{\ps}^*_{\m\a}\g^{\n\r}\mathfrak{C}_{\n\r}-2\bar\xi_\a\g^\n h^*_{\m\n}\right)_{\g\text{-}t}
\left(\g^\m\xi^\a\right)_{\g\text{-}t},\eeq{acc2.5}
where the subscript ``$\g\text{-}t$'' means a $\g$-traceless projection w.r.t.~the $\a$-index. One can follow some steps
similar to those of Eqs.~(\ref{ABdefined}) through~(\ref{2nd8}) to rewrite $c_1^{(\m\b)}$ as:
\bea c_1^{(\m\b)}&=&2\bar\ps^*_{\m\a}[\g^\m,\g^{\n\r}]\mathfrak{C}_{\n\r}\xi^\a-\tfrac{2}{3}\bar\ps^*_{\m\a}
\left(\g^\m\g^\a\g^\b\g^{\n\r}-\g^{\n\r}\g^\a\g^\b\g^\m\right)\xi_\b\mathfrak{C}_{\n\r}\nonumber\\
&&+4\bar\xi_\a\xi^\a h^{*\prime}-\tfrac{16}{3}\bar\xi^\m\xi^\n h^*_{\m\n}.\eea{acc2.6}
The terms containing the graviton antifield $h^*_{\m\n}$ in the second line of the above equation both
vanish by virtue of the Majorana properties of the fermionic ghost $\xi_\a$. In the first line, the first term
calls for the identity~(\ref{commutat1}), whereas the second term simplifies on account of the same
as well as the Clifford algebra, thanks to the $\g$-tracelessness of the fermionic antifield and ghost. The result is:
\beq c_1^{(\m\b)}=8\bar\ps^{*\m\a}\g^\n\xi_\a\mathfrak{C}_{\m\n}-\tfrac{16}{3}\bar\ps^*_{\m\a}
\left(\h^{\m\a}\h^{\n\b}+\h^{\m\b}\h^{\n\a}\right)\g^\r\xi_\b\mathfrak{C}_{\n\r}.\eeq{acc2.7}
Now, by using Eq.~(\ref{g1.9}) one can rewrite the bosonic ghost-curl $\mathfrak{C}_{\m\n}$ to arrive at:
\beq c_1^{(\m\b)}=16\bar\ps^{*\m\a}\left(\de_\m\Cs\right)\xi_\a-\tfrac{32}{3}\bar\ps^*_{\m\a}\left(\h^{\m\a}\h^{\n\b}
+\h^{\m\b}\h^{\n\a}\right)\left(\de_\n\Cs\right)\xi_\b+\G\text{-exact}.\eeq{acc2.8}
Finally, one can get rid of the derivative on the bosonic ghost to write:
\beq c_1^{(\m\b)}\doteq\tfrac{64}{3}\,\de_{(\m}\bar\ps^*_{\n)}{}^\m_{s\text{-}t}\,\Cs\,\xi^\n
+\G\text{-exact}+\D\text{-exact},\eeq{aacc3}
where the subscript ``$s\text{-}t$'' stands for a symmetric-traceless projection. This completes our derivation
of $c_1$ appearing in Eqs.~(\ref{cc2}).

Next, in order to compute $c_0$ we note that Eq.~(\ref{new2.3}) simplifies to:
\beq \int c_0=-\left(\frac{\d^R}{\d\Phi^A}\int a_0\right)\left(\frac{\d^L}{\d\Phi^*_A}\int a_1\right).\eeq{acc1x}
After a straightforward calculation, one finds a $c_0$ of the form:
\beq c_0=c_0^H+\l\a c_0^{(\l\a)}+\a\b c_0^{(\a\b)}+\m^2 c_0^{(\m\m)}+\m\b c_0^{(\m\b)},\eeq{acc2x}
where the nontrivial contributions from the tadpole terms include:
\beq c_0^{(\l\a)}=2C_\n\left(\de^\n h'-2\de_\m h^{\m\n}\right),\quad c_0^{(\a\b)}=-8\bar\xi_\a\pss^\a,
\quad c_0^{(\m\m)}=\left(\bar\ps_{\m\a}\g^{\m\n}\right)_{\g\text{-}t}\left(\g_\n\xi^\a\right)_{\g\text{-}t},
\eeq{aacc4}
plus the more complicated looking $c_0^{(\m\b)}$, given by:
\beq c_0^{(\m\b)}=-2\left(\bar\ps_{\m\a}\g^{\m\n\r}\g^{\s\l}\mathfrak{h}_{\s\l\Vert\r}\right)_{\g\text{-}t}
\left(\g_\n\xi^\a\right)_{\g\text{-}t}-2\left(\bar\ps_{\m\a}\g^{\m\n}\right)_{\g\text{-}t}
\left[\g^{\r\s}\left(\xi^\a\mathfrak{h}_{\r\s\Vert\n}+\ps_\n{}^\a\mathfrak{C}_{\r\s}\right)\right]_{\g\text{-}t}.
\eeq{acc7}
The contributions appearing in Eqs.~(\ref{aacc4}) are rather easy to deal with. We find that
\bea c_0^{(\l\a)}&=&2C_\n\left(\de^\n h'-2\de_\m h^{\m\n}\right)~\doteq~\G\left(h_{\m\n}^2
-\tfrac{1}{2}h^{\prime 2}\right),\label{acc4}\\c_0^{(\a\b)}&=&-8\bar\xi_\a\pss^\a
~\in~H(\G|\D|d).\eea{acc5}
For $c_0^{(\m\m)}$ we can follow some steps similar to those of Eqs.~(\ref{ABdefined})--(\ref{2nd8}) and obtain:
\beq c_0^{(\m\m)}~=~2\bar\pss_\a\xi^\a-\tfrac{1}{3}\ps_{\m\a}[\g^{\m\n}, \g^\a]\{\g^\b, \g_\n\}\xi_\b,\eeq{acc8}
where have used of the $\g$-trace properties of the fermionic field and its ghost. Upon using suitable
$\g$-matrix identities in the second term on the right hand side, we arrive at:
\beq c_0^{(\m\m)}~=~2\bar\pss_\a\xi^\a-\tfrac{4}{3}\bar\pss_\a\xi^\a~=~\tfrac{2}{3}\bar\xi_\a\pss^\a
~\in~H(\G|\D|d).\eeq{acc9}

We are now left with the complicated term $c_0^{(\m\b)}$ given by~(\ref{acc7}). To simplify this term, again
we follow the steps of Eqs.~(\ref{ABdefined})--(\ref{2nd8}) and write the schematic expression:
\beq c_0^{(\m\b)}=-2\left(\mathcal{X}-\tfrac{1}{3}\mathcal{Y}\right),\eeq{acc10}
where the quantities $\mathcal{X}$ and $\mathcal{Y}$ are given by:
\begin{equation}\label{acc11}
\begin{split}
\mathcal{X}&=\left(\bar\ps_{\m\a}\g^{\m\n\r}\g^{\s\l}\mathfrak{h}_{\s\l\Vert\r}\right)
\left(\g_\n\xi^\a\right)+\left(\bar\ps_{\m\a}\g^{\m\n}\right)
\g^{\r\s}\left(\xi^\a\mathfrak{h}_{\r\s\Vert\n}+\ps_\n{}^\a\mathfrak{C}_{\r\s}\right),\\
\mathcal{Y}&=\left(\bar\ps_{\m\a}\g^{\m\n\r}\g^{\s\l}\g^\a\mathfrak{h}_{\s\l\Vert\r}\right)
\left(\g_\b\g_\n\xi^\b\right)+\left(\bar\ps_{\m\a}\g^{\m\n}\g^\a\right)
\g^\b\g^{\r\s}\left(\xi_\b\mathfrak{h}_{\r\s\Vert\n}+\ps_\n{}_\b\mathfrak{C}_{\r\s}\right).
\end{split}
\end{equation}
The first term in $\mathcal{X}$ can be simplified by using identity~(\ref{commutat5}) and then~(\ref{2gamma1}).
In the first term in $\mathcal{Y}$ as well, we can use identity~(\ref{commutat5}) to move $\g^{\m\n\r}$ past
$\g^{\s\l}\g^\a$. Then we can write:
\begin{equation}\label{acc12}
\begin{split}
\mathcal{X}&=\bar\ps_{\m\a}\left(\g^{\m\n}\g^{\r\s}-\g^{\r\s}\g^{\m\n}\right)\xi^\a\mathfrak{h}_{\r\s\Vert\n}
+\bar\ps_{\m\a}\g^{\m\n}\g^{\r\s}\ps_\n{}^\a\mathfrak{C}_{\r\s},\\
\mathcal{Y}&=\bar\ps_{\m\a}[\g^{\s\l},\g^\a]\g^{\m\n\r}\mathfrak{h}_{\s\l\Vert\r}
\{\g_\b,\g_\n\}\xi^\b+\bar\ps_{\m\a}[\g^{\m\n}, \g^\a]
[\g^\b, \g^{\r\s}]\left(\xi_\b\mathfrak{h}_{\r\s\Vert\n}+\ps_\n{}_\b\mathfrak{C}_{\r\s}\right),
\end{split}
\end{equation}
where in the second line we have made use of the $\g$-tracelessness of the fermionic field and ghost.
These expressions simplify on account of identities~(\ref{commutat1}) and~(\ref{commutat2}), giving:
\begin{equation}\label{acc13}
\begin{split}
\mathcal{X}&=4\bar\ps_{\m\a}\left(\h^{\n\r}\g^{\m\s}-\h^{\m\r}\g^{\n\s}\right)\xi^\a\mathfrak{h}_{\r\s\Vert\n}
+2\bar\ps_{\m\a}\left(\g^\m\h^{\n\r}-\g^\n\h^{\m\r}\right)\g^\s\ps_\n{}^\a\mathfrak{C}_{\r\s},\\
\mathcal{Y}&=-8\bar\ps_\m{}^\s\g^\l\g^{\m\n\r}\xi_\n\mathfrak{h}_{\s\l\Vert\r}+8\left(\bar\pss^\n
-\bar\ps'\g^\n\right)\g^\s\left(\xi^\r\mathfrak{h}_{\r\s\Vert\n}+\ps_\n{}^\r\mathfrak{C}_{\r\s}\right),
\end{split}
\end{equation}
where in the first line we have dropped the term $\bar\ps_{\m\a}\ps_\n{}^\a\mathfrak{C}^{\m\n}$, which vanishes
identically. Because of the $\g$-tracelessness of $\xi_\n$ the first term in $\mathcal{Y}$ can be further
simplified, by using the expansion~(\ref{3gamma1.5}), to the form:
$8\bar\ps_\m{}^\s\g^\r\left(\g^\m\xi^\n-\g^\n\xi^\m\right)\mathfrak{h}_{\r\s\Vert\n}$.

The next step is to make the graviton field $h_{\m\n}$ undifferentiated. Then this derivative
hits either on the Majorana-conjugated fermion $\bar\ps_{\m\a}$ (and its $\g$-traces) or on the ghost field $\xi^\a$.
It is not difficult to see that the derivative terms of $\bar\ps_{\m\a}$ comprise $\D$-exact pieces individually
in $\mathcal{X}$ and $\mathcal{Y}$, thanks to the various forms of the fermion EoMs~(\ref{eomsmagic})--(\ref{eomsgtr2.6}).
It is a bit of an exercise to show that in the remaining terms containing $h_{\m\n}$, the fermion bilinears combine into
$\G$-exact pieces. Various properties of the Majorana fermions $\ps_{\m\a}$ and $\xi_\a$ play important roles in showing
this. The result is:
\begin{equation}\label{acc14}
\begin{split}
\mathcal{X}&\doteq4\left(h_{\m\n}-\tfrac{1}{2}\h_{\m\n}h'\right)\G\left(\bar\ps_{\r\a}\g^{\r\m}\ps^{\n\a}\right)
-4\left(\bar\ps_{\r\a}\g^{\r\m}\ps^{\n\a}\right)\mathfrak{C}_{\m\n}+\D\text{-exact},\\
\mathcal{Y}&\doteq h_{\r\s}\,\G\left[\left(\bar\pss^\n-\bar\ps'\g^\n\right)\g^\s\ps_\n{}^\r-2\bar\ps^{\r\n}\ps_\n{}^\s
+2\bar\ps'\ps^{\r\s}-\bar\ps_\m{}^\r\g^{\m\n}\ps_\n{}^\s\right]\\
&+4h'\,\G\left(\bar\ps_\m{}^\n\ps_\n{}^\m -\bar\ps'\ps'\right)
+8\left(\bar\pss^\n-\bar\ps'\g^\n\right)\g^\s\ps_\n{}^\r\mathfrak{C}_{\r\s}+\D\text{-exact}.
\end{split}
\end{equation}
In the above expressions we can pick up $\G$-exact terms at the cost of introducing a (symmetrized)
derivative of the bosonic ghost $C_\m$. The latter kind of terms then simplify against the already existing terms
containing the ghost-curl $\mathfrak{C}_{\m\n}$. Then we can make the bosonic ghost undifferentiated,
which gives us the following form:
\beq \mathcal{X}\doteq 8 C_\r\de_\s\mathcal{X}^{\r\s}+\G\text{-exact}+\D\text{-exact},\qquad
\mathcal{Y}\doteq 16 C_\r\de_\s\mathcal{Y}^{\r\s}+\G\text{-exact}+\D\text{-exact}\eeq{acc15}
where $\mathcal{X}^{\r\s}$ and $\mathcal{Y}^{\r\s}$ are the following fermion bilinears:
\begin{equation}\nonumber
\begin{split}
\mathcal{X}^{\r\s}&=\bar\ps_{\m\a}\g^{\m\s}\ps^{\r\a}-\tfrac{1}{2}\h^{\r\s}\bar\ps_{\m\a}\g^{\m\n}\ps_\n{}^\a,\\
\mathcal{Y}^{\r\s}&=\bar\ps_{\m\n}\g^{\m\s}\ps^{\n\r}-\bar\ps'\g^{\m\s}\ps_\m{}^\r-\bar\ps_\m{}^\r\g^{\m\n}\ps_\n{}^\s
-\bar\ps^{\r\m}\ps_\m{}^\s+\bar\ps^{\r\s}\ps'+\tfrac{1}{2}\h^{\r\s}\left(\bar\pss^\n-\bar\ps'\g^\n\right).
\end{split}
\end{equation}
From these explicit expressions it is easy to see that the quantities $\de_\s\mathcal{X}^{\r\s}$ and
$\de_\s\mathcal{Y}^{\r\s}$ are $\D$-exact, thanks again to the EoMs~(\ref{eomsmagic})--(\ref{eomsgtr2.6}).
This means from Eq.~(\ref{acc15}) that both $\mathcal{X}$ and $\mathcal{Y}$ are trivial elements in $H(\G|\D|d)$,
so that Eq.~(\ref{acc10}) gives:
\beq c_0^{(\m\b)}\doteq \G\text{-exact}+\D\text{-exact}.\eeq{acc16}

Now, let us plug all the results~(\ref{acc4}),~(\ref{acc5}),~(\ref{acc9}) and~(\ref{acc16}) into
the schematic expression~(\ref{acc2x}). Then we obtain the desired $c_0$ given in Eqs.~(\ref{cc2}):
\beq c_0\doteq c_0^H+\left(\tfrac{2}{3}\m^2-8\a\b\right)\bar\xi_\a\pss^\a+\G\text{-exact}+\D\text{-exact}.
\eeq{acc2xx}

Finally, we look at the computation of $c_2$, $c_1$ and $c_0$ in Section~\ref{subsec:CC2}.
While $c_2$ will be the same as that of
Section~\ref{sec:GenH}, $c_1$ and $c_0$ can be obtained by closely inspecting the steps of derivation
of their spin-5/2 counterparts. Note that the rank $n$ plays a nontrivial role only in defining the
$\g$-traceless projections (compare Eqs.~(\ref{2nd8}) and~(\ref{g2nd8}), for example). Apart from
straightforward proliferation of indices, the generalization $s=5/2\rightarrow(n+3/2)$ essentially
boils down to the factor replacement: $\tfrac{1}{3}\rightarrow\left(\tfrac{n}{2n+1}\right)$ in front of those terms
that originate solely from $\g$-traceless projections. Examples of the latter kind are the second and fourth terms
of Eq.~(\ref{acc2.6}), and the second terms in Eqs.~(\ref{acc8}) and~(\ref{acc10}) both. Following the subsequent
steps in the respective computations, it is easy to see that the arbitrary-spin generalizations of Eqs.~(\ref{cc2})
will be given by Eqs.~(\ref{gcc2}).

\end{appendix}

\end{document}